\documentclass[useAMS,usenatbib]{style/mnras}

\usepackage{epsfig,amsmath,natbib,wasysym}
\usepackage{color}

\usepackage{pifont}
\usepackage{xcolor}
\usepackage{amssymb}
\usepackage{multirow}
\usepackage{framed}
\usepackage{subcaption} 

\usepackage{placeins} 

\usepackage{etoolbox}
\makeatletter
\newcount\c@additionalboxlevel
\newcount\c@maxboxlevel
\patchcmd\@combinedblfloats{\box\@outputbox}{%
  \stepcounter{additionalboxlevel}%
  \box\@outputbox
}{}{\errmessage{\noexpand\@combinedblfloats could not be patched}}

\AtBeginShipout{%
  \ifnum\value{additionalboxlevel}>\value{maxboxlevel}%
    \typeout{Warning: maxboxlevel might be too small, increase to %
      \the\value{additionalboxlevel}%
    }%
  \fi 
  \@whilenum\value{additionalboxlevel}<\value{maxboxlevel}\do{%
    \typeout{* Additional boxing of page `\thepage'}%
    \setbox\AtBeginShipoutBox=\hbox{\copy\AtBeginShipoutBox}%
    \stepcounter{additionalboxlevel}%
  }%
  \setcounter{additionalboxlevel}{0}%
}
\makeatother

\def\refjnl#1{{\rm#1}}
\def\apjl{\refjnl{ApJ}}                
\def\apjs{\refjnl{ApJS}}               
\def\ao{\refjnl{Appl.~Opt.}}           
\def\aap{\refjnl{A\&A}}                

\def\refjnl#1{{\rm#1}}
\def\apjl{\refjnl{ApJ}}                
\def\apjs{\refjnl{ApJS}}               
\def\ao{\refjnl{Appl.~Opt.}}           
\def\aap{\refjnl{A\&A}}                

\def\be{\begin{equation}} 
\def\ee{\end{equation}} 
\def\ba{\begin{eqnarray}} 
\def\ea{\end{eqnarray}}

\def\kms{\,{\rm {km\, s^{-1}}}} 
\def\cc{\,{\rm {cm^{-3}}}} 
\def\msun{{\Msun}} 
\def\lsun{{\Lsun}}

  \def\erg{{\rm erg}}

\def\gsim{\lower.5ex\hbox{\gtsima}} 
\def\lsim{\lower.5ex\hbox{\ltsima}} \def\gtsima{$\; \buildrel > \over 
\sim \;$} \def\ltsima{$\; \buildrel < \over \sim \;$} \def\prosima{$\; 
\buildrel \propto \over \sim \;$} \def\gsim{\lower.5ex\hbox{\gtsima}} 
\def\lsim{\lower.5ex\hbox{\ltsima}} 
\def\simgt{\lower.5ex\hbox{\gtsima}} 
\def\simlt{\lower.5ex\hbox{\ltsima}} 
\def\simpr{\lower.5ex\hbox{\prosima}} \def\la{\lsim} \def\ga{\gsim} 
  
 \def\gtsima{$\; \buildrel > \over \sim \;$} 
\def\ltsima{$\; \buildrel < \over \sim \;$} 
\def\gsim{\lower.5ex\hbox{\gtsima}} 
\def\lsim{\lower.5ex\hbox{\ltsima}} 
\def\simgt{\lower.5ex\hbox{\gtsima}} 
\def\simlt{\lower.5ex\hbox{\ltsima}} 
\def\simpr{\lower.5ex\hbox{\prosima}} 
\def\la{\lsim} 
\def\ga{\gsim}

\def\msun{\,{\rm \Msun}}

\def\E3{{\cal E}_{\rm g}^{III}}

\def\Msun{M_\odot}

\def\x12{x_{1/2}} 
\def\v12{v_{1/2}}

\newcommand{\quotes}[1]{``#1''}

\def\ramses{$\textsc{ramses}$}
\def\ramsesrt{$\textsc{ramses-rt}$}
\def\krome{$\textsc{krome}$}
\def\cloudy{$\textsc{cloudy}$}
\def\pymses{$\textsc{pymses}$}
\def\parsec{$\textsc{parsec}$}

\def\yr{{\rm yr}}
\def\lsun{{\rm L}_{\odot}}
\def\msun{{\rm M}_{\odot}}

\def\h2{{\mathrm{H}_2}}
\def\hii{\mathrm{HII}}
\def\hi{\mathrm{HI}}

\def\kms{\mathrm{km}\,\mathrm{s}^{-1}}
\def\ergs{\mathrm{erg}\,\mathrm{s}^{-1}}
\def\gmc{\textsc{gmc}}
\def\ism{\textsc{ism}}
\def\sfr{\mathrm{SFR}}
\def\sfrd{\mathrm{SFRd}}

\definecolor{editcolor}{HTML}{34bf1f}

\title[Shaping the structure of a GMC with radiation and winds]{Shaping the structure of a GMC with radiation and winds}
\author[Decataldo et al.]{
D. Decataldo$^{1}$\thanks{\href{mailto:davide.decataldo@sns.it}{davide.decataldo@sns.it}},
A. Lupi$^{1}$,
A. Ferrara$^{1}$, 
A. Pallottini$^{1,2}$, 
and M. Fumagalli$^{3,4,5}$
\\
$^{1}$ Scuola Normale Superiore, Piazza dei Cavalieri 7, I-56126 Pisa, Italy\\
$^{2}$ Centro Fermi, Museo Storico della Fisica e Centro Studi e Ricerche \quotes{Enrico Fermi}, Piazza del Viminale 1, Roma, 00184, Italy\\
$^{3}$ Dipartimento di Fisica G. Occhialini, Universit\`a degli Studi di Milano Bicocca, Piazza della Scienza 3, 20126 Milano, Italy \\
$^{4}$ Institute for Computational Cosmology, Durham University, South Road, Durham, DH1 3LE, UK \\
$^{5}$ Centre for Extragalactic Astronomy, Durham University, South Road, Durham, DH1 3LE, UK \\
}

\begin{document}
 
\date{\today} 
 
\pagerange{\pageref{firstpage}--\pageref{lastpage}} \pubyear{2020}
 
\maketitle 

\setlength{\parskip}{0pt}
    
\label{firstpage}
\begin{abstract}
We study the effect of stellar feedback (photodissociation/ionization, radiation pressure and winds) on the evolution of a Giant Molecular Cloud (GMC), by means of a 3D radiative transfer, hydro-simulation implementing a complex chemical network featuring $\h2$ formation and destruction. We track the formation of individual stars with mass $M>1\,\msun$ with a stochastic recipe. Each star emits radiation according to its spectrum, 
sampled with 10 photon bins from near-infrared to extreme ultra-violet bands; winds are implemented by energy injection in the neighbouring cells. We run a simulation of a GMC with mass $M=10^5\,\msun$, following the evolution of different gas phases. Thanks to the simultaneous inclusion of different stellar feedback mechanisms, we identify two stages in the cloud evolution: (1) radiation and winds carve ionized, low-density bubbles around massive stars, while FUV radiation dissociates most $\h2$ in the cloud, apart from dense, self-shielded clumps; (2) rapid star formation (SFR$\simeq 0.1\,\msun\,\yr^{-1}$) consumes molecular gas in the dense clumps, so that UV radiation escapes and ionizes the remaining $\hi$ gas in the GMC. $\h2$ is exhausted in $1.6$ Myr, yielding a final star formation efficiency of 36 per cent. 
The average intensity of FUV and ionizing fields increases almost steadily with time; by the end of the simulation ($t=2.5$ Myr) we find $\langle G_0 \rangle \simeq 10^3$ (in Habing units), and a ionization parameter $\langle U_{\rm ion} \rangle \simeq 10^2$, respectively. The ionization field has also a more patchy distribution than the FUV one within the GMC. Throughout the evolution, the escape fraction of ionizing photons from the cloud is $f_{\rm ion, esc} \simlt 0.03$.
\end{abstract}



\section{Introduction}\label{introduction}

The structure of giant molecular clouds (GMC) is determined by the effect of different phenomena. Initial turbulence can be be established during the formation process, possibly inherited from large scale turbulence present in the arms of the parent galaxy \citep{Brunt2009, Elmegreen2011a, Hughes2013, Colombo2014, Dobbs2015, Walch2015}. Then gravity kicks in, fostering the formation of dense clumps and filaments, and at the same time leaving behind low density regions devoid of gas. The gas keeps collapsing in overdense regions, until the density is high enough for star formation to occur. Stars with larger masses (above 10 $\msun$) have a dramatic impact over the surrounding interstellar medium (ISM), interacting with it through different mechanisms: radiation \citep{Whitworth1979}, winds \citep{Castor1975, Weaver1977} and supernova explosions \citep{Sedov1958, Ostriker1988}. Hence, the effect of stellar feedback is crucial in determining the structure of GMCs, and a proper modelling is needed in order to understand their evolution.

Stellar feedback is indeed invoked to explain many of the observed cloud properties. For example, observations show that GMCs in our Galaxy have in general a very low star formation efficiency (SFE) and star formation rate (SFR), converting only a few percent of the gas mass to stars \citep{Williams1997, Carpenter2000, Krumholz2005a, Evans2009, Garcia2014, Lee2016, Chevance2019}. A possible explanation is that GMCs are indeed short-lived, and gas is quickly heated and dispersed by stellar feedback, therefore quenching completely star formation in less then 10 Myr \citep{Elmegreen2000, Hartmann2001}. This picture is supported by the recent analysis of the CO-to-H$\alpha$ ratio, which is an indicator of the co-spatiality of the GMC molecular phase and young stars \citep{Kruijssen2019, Chevance2019}. On the other side, the existence of long-lived GMCs would require stellar feedback to increase the turbulence in order to provide pressure support, without dispersing the cloud \citep{Federrath2012, Padoan2012}.

High-mass stars ($M>5\,\msun$) emit an important fraction of photons in the far ultraviolet (FUV, $6\,{\rm eV}<h\nu < 13.6\,{\rm eV}$) and extreme ultraviolet bands (EUV, $h\nu > 13.6\,{\rm eV}$). FUV photons can dissociate molecules like CO and $\h2$, determining the formation of molecular-to-atomic transition regions (photodissociation regions, PDR) heated up to $10^3$ K \citep{Tielens1985, Kaufman1999, LePetit2006, Bron2018}; EUV photons ionize hydrogen and helium ($\hii$ regions), rising the gas temperatures to $10^{4-5}$ K depending on the luminosity of the star \citep{Stromgren1939, Anderson2009}. Since hot gas has higher pressure relative to the cold molecular ISM, shocks propagate ahead of dissociation/ionization fronts, compressing the gas and driving turbulent motions \citep{Kahn1954, Williams2018}. Gas photoevaporates from dense clumps, hence reducing their molecular mass, but at the same time the radiation-driven implosion can trigger star formation \citep{Kessel-Deynet2003, Bisbas2011, Decataldo2019}.

Stars larger then $10-12\,\msun$ eject mass with a rate of about $10^{-7}-10^{-5} \, \msun\,\yr^{-1}$, with terminal velocities up to $3000\,\kms$ \citep{Leitherer1992}. The ejected gas shocks and sweeps away the surrounding medium, leaving a hot low density bubble around the star, with temperatures as high as $T\sim 10^{6-7}$ K \citep{Weaver1977,McKee1977,Cioffi1988,Ostriker1988}. Both radiation and winds keep injecting energy in the ISM for the whole lifetime of the star, hence resulting in a large energy input ($10^{50-52}$ erg for a $10^6\,\lsun$ star living about 3 Myr). However, it is still unclear which of the two feedback mechanisms is more effective, due to the different coupling efficiency with the gas \citep{Matzner2002, Walch2012, Haid2018}. Supernova explosions occur at the end of stellar life, releasing about $10^{51}$ erg in a very short time. The coupling of the supernova blast with the cloud gas strongly depends on the density structure, since dense gas cools efficiently radiating away all the energy input \citep{Thornton1998, Dwarkadas2012, Walch2015b}. 

Pioneer analytical works \citep{Whitworth1979, Williams1997, Matzner2002} have analysed the problem of an HII region expanding in a molecular cloud, estimating the amount of ionized/dispersed gas. EUV radiation has been identified as the main responsible for cloud destruction and hence star formation inefficiency. In order to account for the complex structure of a realistic GMC, numerical simulations have been recently employed to study their evolution under the effect of stellar feedback. Early simulations have used approximated recipes to account for stellar photoionization, without the inclusion of radiative transfer calculations: for example, a common approach is to compute the ionization state of the gas and then the corresponding temperature \citep{Dale2005, Dale2007a, Ceverino2009, Gritschneder2009}, or to account for photoionization feedback by injecting thermal energy at stellar locations \citep{Vazquez-Semadeni2010}. Nevertheless, these models managed to reproduce the main features of typical observed structures, as pillars and bright-rimmed clumps, and they assess the effect of triggered star formation due to radiation feedback \citep[see review by][]{Elmegreen2011}.

More recent works have adopted coupled hydrodynamics and radiative transfer schemes to account for gas-radiation interaction.
\citet{Walch2012} study the effect of radiation from one massive star placed at the centre of a fractal $10^4\,\msun$ GMC, getting a speed up of star formation with respect to a control case with no radiation, while the overall star formation efficiency (SFE=$M_{\star}/M_{\rm GMC}$, i.e. the ratio between the stellar mass and the initial GMC mass) is reduced.
\citet{Raskutti2016} follow the evolution of many GMCs with different mass, radius (hence different gas surface density $\Sigma$) and initial amount of turbulence. Star formation is implemented via sink particles (each representing a stellar cluster), which emit radiation with intensity proportional to their mass. They find SFEs of the order of $0.1-0.6$, increasing proportionally to $\log \Sigma$; for all the clouds most of stars form in a time shorter than one free-fall time $t_{\rm ff}$. They define the lifetime of a cloud as the time $t_\ell$ at which the virial parameter $\alpha=5$, corresponding to a sheer drop of the star formation rate, yielding $t_\ell/t_{\rm ff} \sim 1.2-1.9$.

A similar analysis is carried out by \citet{Howard2017}, finding that radiation feedback reduces the SFE more consistently for more massive clouds ($10^4-10^6\,\Msun$) with respect to smaller clouds ($\sim10^3\,\Msun$), since massive clouds manage to form a richer population of massive stars. In their simulations ${\rm SFE}\sim 0.3-0.6$, suggesting that radiation feedback alone is not enough to suppress the SFE to the generally observed values \citep[$\sim 1-10\%$][]{Lada2010, Murray2010, Lee2016, Lada2016, Ochsendorf2017}. However, other models show that a lower SFEs can be obtained by including magnetic fields in the simulation \citep{Geen2016, Geen2018, Kim2018, Haid2019, He2019}. 
In our paper, we make a further step in the accuracy of GMC simulations, including several novel features:
\begin{itemize}
    \item multi-bin radiative transfer, sampling radiation in the near-infrared, far-ultraviolet and extreme-ultraviolet bands;
    \item non-equilibrium chemical network coupled with the radiative transfer scheme;
    \item simultaneous inclusion of different stellar feedback mechanisms: photo-ionization/dissociation, radiation pressure, winds, supernovae;
    \item stellar particles representing individual stars and not star clusters, each one emitting radiation with a spectrum derived from stellar tracks.
\end{itemize}
The goal is to understand how stellar feedback alters the structure and the chemical composition of a GMC, determining its final star formation efficiency and lifetime.

The paper is organised as follows. In Sec. \ref{numerical_simulation}, we describe the numerical method and the simulations suite, in particular the inclusion of radiation (Sec. \ref{sec:sub:model_radiation}), winds (Sec. \ref{sec:sub:model_winds}) and SN (Sec. \ref{sec:sub:model_SN}). In Sec. \ref{results}, snapshots of the clouds are shown (Sec. \ref{subsec:overview} and \ref{subsec:hii_regions}), and its features are analysed with time, focusing on the evolution of the ionized, atomic and molecular phases (Sec. \ref{subsec:ism_phases}), the star formation efficiency and the star formation rate (Sec. \ref{subsec:sfe_and_sfr}) and the radiation field (Sec. \ref{subsec:radiation}). Our conclusions are finally summarised in Sec. \ref{sec:conclusions}.


\section{Numerical Simulation}
\label{numerical_simulation}

\begin{table}
\centering
\begin{tabular}{ll}
\multicolumn{2}{c}{Photoreactions}{$E_{\rm act}$}                                                                                            \\ \hline
${\rm H}+\gamma \rightarrow {\rm H}^+ + e$         & 13.6 eV\\
${\rm H}_2^+ +\gamma \rightarrow {\rm H}^+ + {\rm H}$ & 2.65 eV       \\
${\rm He}+\gamma \rightarrow {\rm He}^+ + e$       & 24.6 eV \\
${\rm H}_2^+ +\gamma \rightarrow {\rm H}^+ + {\rm H}^+ + e$ &  30.0 eV\\
${\rm He}^+ +\gamma \rightarrow {\rm He}^{++} + e$ & 54.4 eV \\
${\rm H}_2 +\gamma \rightarrow {\rm H} + {\rm H}\,\,$ (direct) & 14.2 eV \\
${\rm H}^- +\gamma \rightarrow {\rm H} + e$        & 0.76 eV\\
${\rm H}_2 +\gamma \rightarrow {\rm H} + {\rm H}\,\,$ (Solomon) & 11.2 eV\\
${\rm H}_2 +\gamma \rightarrow {\rm H}_2^+ + e$    & 15.4 eV \\                                                                     
\end{tabular}
\caption{List of photochemical reactions included in our chemical network, and corresponding activation energy.
\label{photon_bins}
}
\end{table}

\begin{figure}
\centering
\includegraphics[width=0.49\textwidth]{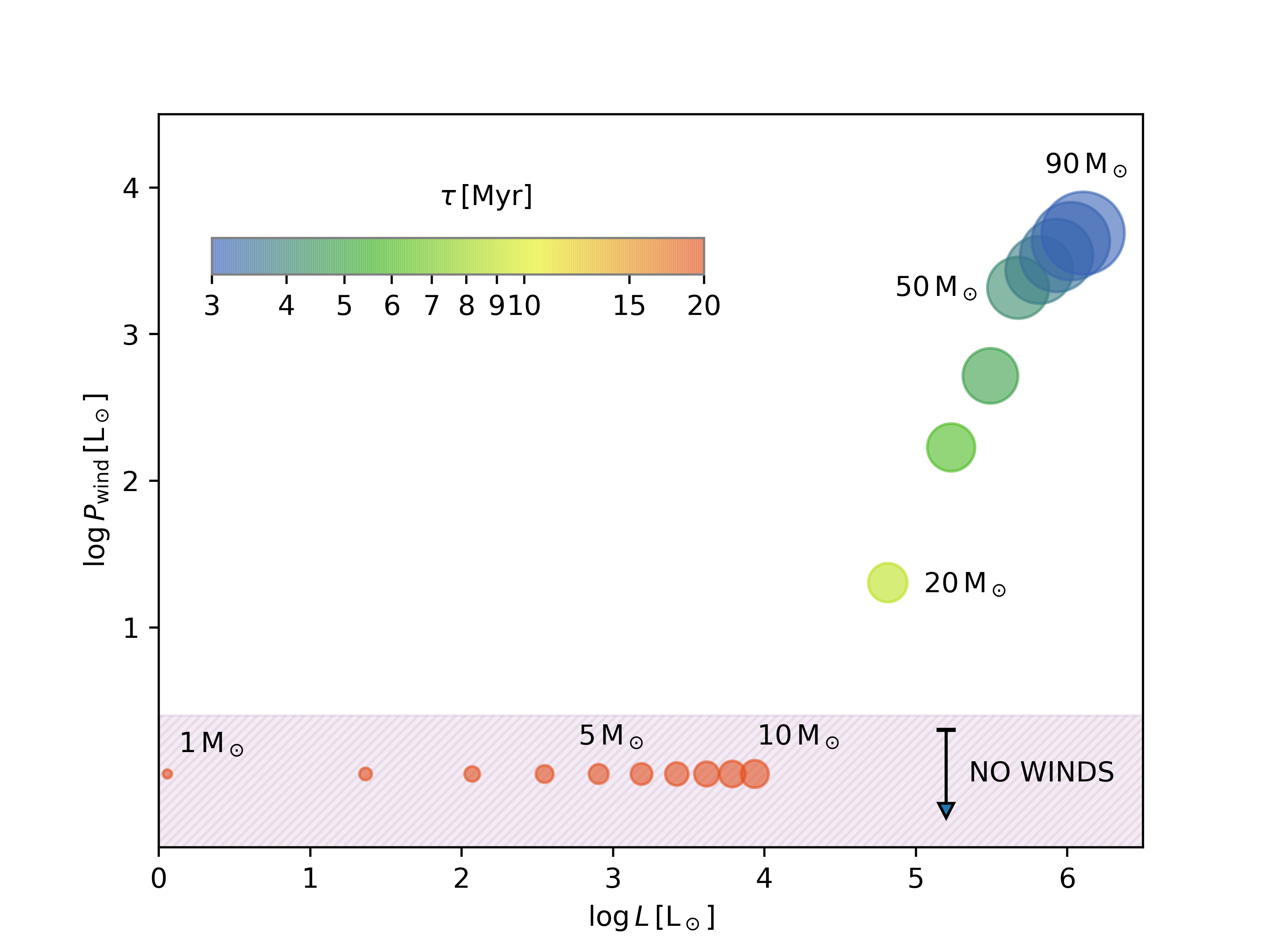}
\caption{Overview of properties of stars on the main sequence. Circles represent stars of different masses, with the diameter of the circle proportional to the mass. The color of each circle corresponds to the time interval (also encoded in the colorbar) a star spends on the main sequence, with red circles standing for times longer than 20 Myr. On the {\it x} and {\it y} axis the bolometric luminosity and wind kinetic power are shown, both expressed in units of solar luminosities. Stars with mass $<12 \msun$ do not produce significant winds.  
\label{star_properties}}
\end{figure}

\begin{figure}
\centering
\includegraphics[width=0.5\textwidth]{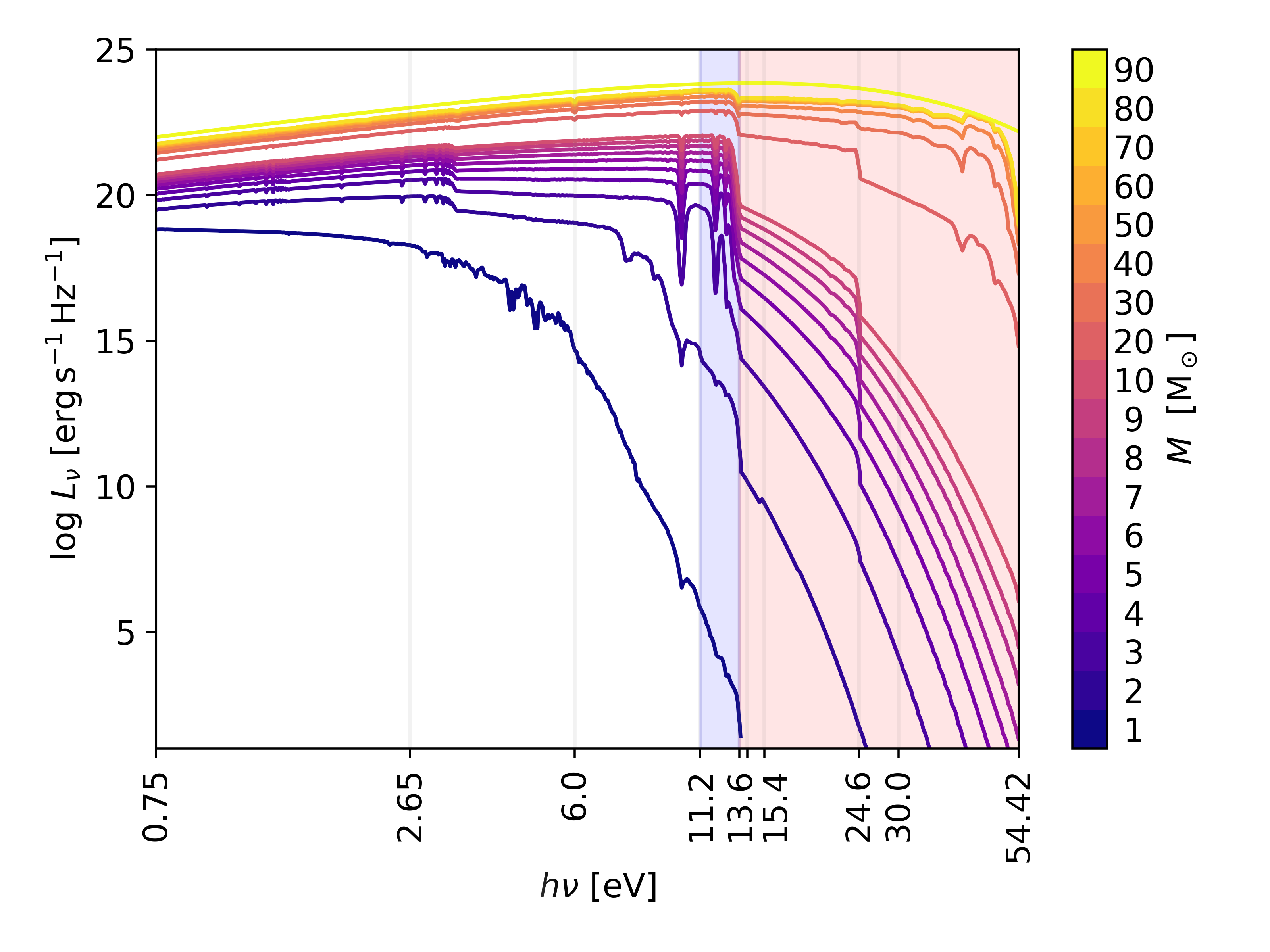}
\caption {Spectra of stars with different masses, the same considered in Fig. \ref{star_properties}. The vertical grey lines delimit the energy bins adopted in the our simulation suit. The blue and red areas highlight the LW and the EUV bands, respectively.
\label{star_spectra}
}
\end{figure}

We carry out our simulations using a customised version of the adaptive mesh refinement (AMR) code \ramsesrt~\citep{Teyssier2002, Rosdahl2013}. The basic version \ramses~features a second-order Godunov scheme for the gas hydro-dynamics and a particle-mesh solver for particles as stars. \ramsesrt~implements radiative transfer (RT) via a momentum-based approach, using a first-order Godunov solver and the M1 closure relation for the Eddington tensor. In order to have reasonable timesteps in the simulation, we adopt a reduced speed of light $c_{\rm red}=10^{-3}\,c$, where $c$ is the physical speed of light. This approximation brings to an inaccurate propagation of the ionization front IF only when its speed $v_{\rm IF}$ is larger then $c_{\rm red}$ \citep{Deparis2019, Ocvirk2019}, which in our simulation happens only close to very massive stars.
The thermochemistry module of \ramsesrt~has been coupled with the package \krome \citep{Grassi2014} in order to include a complex chemical network accounting for 9 species (H, $\mathrm{H}^{+}$, $\mathrm{H}^{-}$, $\mathrm{H}_2$, $\mathrm{H}_2^{+}$, He, $\mathrm{He}^{+}$, $\mathrm{He}^{++}$ and free electrons). This coupling between \ramsesrt~and \krome~has been already been tested and used in previous works \citep{Pallottini2019, Decataldo2019}.

The chemical network consists of 46 reactions in total, taken from the ones listed in \citet{Bovino2016}: reactions 1 to 31, 53, 54 and from 58 to 61 in their tables B.1 and B.2, photoreactions P1 to P9 in their table 2. Photoreactions are also listed in Tab. \ref{photon_bins} for conveniency, with the corresponding activation energies $E_{\rm act}$. We adopted the same rates, with the exception of $\h2$ formation on dust grains.

$\h2$ forms mainly on the surface of dust grains: H atoms stick on the grain and migrate over its surface, then associating with another H atom and finally evaporating from the grain \citep{Jura1975}. An increase of the temperature raises the probability of a collision between H atoms and grains, but also decreases the sticking coefficient. Hence, we adopt a temperature-dependent $\h2$ formation rate, given by \citet{Hollenbach1979} and \citet{Sternberg1989}
\be
R_{\rm f}=3\times 10^{-17}\, f_a \, S \, \left(\dfrac{T}{100 {\,\rm K}}\right)^{0.5} \, \cc\,{\rm s}^{-1}
\ee
where the factor $f_a$ is the fraction of atoms that do not evaporate before forming $\h2$ and $S$ is the gas-grain sticking coefficient:
\be
f_a = \left[1+10^4 \exp\left(-\dfrac{600 {\, \rm K}}{T_d}\right) \right]^{-1}
\ee
\be 
S = \left[ 1 + \left(\dfrac{T+T_d}{625 {\,\rm K}}\right)^{0.5} + \dfrac{T}{500 {\,\rm K}} + 2  \left(\dfrac{T}{500 {\,\rm K}}\right)^2 \right]^{-1}
\ee
In the above expressions, $T$ is the gas temperature and $T_d$ is the dust temperature, which we assume to be $T_d = 30$ K.

We track radiation using 10 energy bins, which have been chosen to cover the energies of interest for the 9 photoreactions included in the chemical network. Radiation is absorbed independently in each bin, taking into account (1) photons taking part in chemical reactions, (2) $\h2$ self-shielding and (3) dust absorption \citep[for details on the implementation, see][]{Decataldo2019}. The adopted $\h2$ self-shielding factor by LW radiation is taken from \citet{Richings2014b}, while opacities for dust absorption are taken from \citet{Weingartner2001}. We have used the Milky Way size distribution for visual extinction-to-reddening ratio $R_V =3.1$, with carbon abundance (per H nucleus) $b_C=60$ ppm in the log-normal populations\footnote{\url{www.astro.princeton.edu/~draine/dust/dustmix.html}}.

The thermal state of the gas is computed by including several heating and cooling mechanisms. At high temperature, the main heating mechanism is photoheating of H and He, computed as in Sec. 2.2 of \citet{Grassi2014}. Photoelectric heating from dust, heating due to exoenergetic reactions, and cosmic ray heating are also included. On the other hand, the gas cools via collisional ionization, collisional excitation and recombination of H and He \citep[rates from][]{Cen1992}, free-free cooling, cooling by $\h2$ (\citealt{Glover2008} and metal cooling. For the metal cooling, we have adopted a cooling function obtained with \cloudy \cite[version 10.00,][]{Ferland1998} for an \citet{Haardt2012} diffuse UV background. We notice that the assumption of an UV background has been commonly adopted also in other simulations of molecular clouds \citep[e.g.][]{Walch2015, Geen2016, Haid2019} to compute the cooling function, but near strong radiation sources the shape of the cooling function is modified due to the different ionization state of metals \citep{Gnedin2012}, hence entailing an error in the equilibrium temperature of ionized gas. For example, in a single-cell test with \krome, we obtain a temperature of $\sim 1.3\times 10^4$ in a cell with gas density $n=100\,\cc$ and the flux of a $30\,\msun$ star at a distance of 1 pc, while a cooling function modified for such flux \citep{Gnedin2012} gives $T\sim 9\times 10^3$ K.

\subsection{Initial conditions}

The GMC  is initially a uniform spherical cloud with mass $M=10^5\,\msun$ and radius $R=20$ pc, implying a number density $n=120\,\cc$ and a free fall time $t_{\rm ff}=\sqrt{3\pi/32G\rho}\simeq 4.7$ Myr. In the MW distribution of GMC properties, these kind of clouds are the most abundant \citep{Heyer2009, Grisdale2018}. The GMC is placed at the centre of a cubic box with size 60 pc, immersed in a uniform background medium with number density $n_\ism=1\,\cc$. The gas has the same chemical composition through all the box, with helium abundance of 25 per cent and hydrogen in fully molecular form. The initial temperature is set to 10 K everywhere.

We add a turbulent velocity field in the initial conditions. We generate an isotropic random Gaussian velocity field with power spectrum $P(k) \propto k^{-4}$ in Fourier space, normalising the velocity perturbation in the following way: inside the GMC, the velocity field is such that the cloud virial parameter $\alpha_\gmc = 5 \,v_{\rm rms}^2 \, R_\gmc / G\,M_\gmc=2$, so that the cloud is initially unbound; in the background medium, the velocity field has a root mean squared value which is $(n_\ism/n_\gmc)^2$ higher than the one in the cloud; this ensures that the ram pressure $P=\rho v^2$ reaches equilibrium at the cloud boundary. In three dimensions, the chosen power spectrum gives a velocity dispersion that varies with scale as $\ell^{1/2}$, i.e. in agreement with Larson's scaling relations \citep{Larson1981}.

The coarse resolution is $2^6$ cells for the background medium (corresponding to a cell size of $\Delta x\simeq 0.9$ pc) and $2^8$ cells for the cloud ($\Delta x\simeq 0.2$ pc). The resolution in the cloud is increased by two further levels of refinement ($\Delta x\simeq 0.06$ pc), according to a Lagrangian strategy: a cell at level $l$ is refined if the gas mass contained exceeds $M_l={\rm k}_l M_\textsc{sph}$, where $M_\textsc{sph}\simeq 10^{-4}\,\msun$, ${\rm k}_9=32$ and ${\rm k}_{10}=24$. In this way, the resolution is increased in denser regions, such as clumps and filaments which are expected to form during the gravitational collapse and due to the effect of stellar feedback.

\subsection{Star formation}\label{star_formation}

We enable star formation in the GMC after 3 Myr (corresponding to $\sim 0.6\,t_{\rm ff}$), when the dense filaments and clumps have formed within the cloud because of gravitational instabilities. Hence, we define the evolutionary time of the cloud $t$ as the time elapsed since $t_0=3$ Myr. Since we are mostly interested in the effect of feedback by UV radiation and stellar winds over the GMC, we neglect the process of gas accretion onto seed particles, and stars are formed directly with their zero-age main sequence mass. For a given cloud mass $M_\gmc$ and a given local star formation efficiency $\eta$, star formation is implemented in the following steps, which will be detailed below:
\begin{itemize}
    \item[(1)] we generate a list of stars before starting the simulation;
    \item[(2)] at each timestep, a star formation episode is triggered in a cell with a probability proportional to the local SFR;
    \item[(3)] the next star in the list is placed in the cell where a star formation episode has occurred.
\end{itemize}
The list includes stars for a total stellar mass equal to the GMC mass ($M_{\rm stars}=M_\textsc{gmc}=10^5\,\msun$), but we do not expect to actually form all the stars in the list, due to the effect of feedback reducing the mass available for star formation. 

The star masses in the list are drawn from a Kroupa Initial Mass Function \citep{Kroupa2001}, until $M_{\rm stars}$ is reached. From this list of stars, we remove the stars with mass lower than 1 $\msun$, since these stars have weak emission in the UV, no winds and they do not explode as supernovae. Hence, we can save computational time by not tracking these stars in the simulation. A factor $f_\star$ is then introduced in the star formation routine, to keep the SFR consistent (see eq. \ref{prob_sf} below).    

Given the list of stars, they are placed in the GMC one by one at runtime, by deciding which cells host a star formation event in the following way. The local star formation rate density ($\sfrd$) is taken to be proportional to the gas density ($\rho$) \citep{Schmidt1959, Kennicutt1998}, which has been shown to hold for clouds in the Milky Way \citep{Krumholz2012}:
\be 
\sfrd = \eta \, \dfrac{\rho}{t_{\rm ff}} \ ,
\ee
where $\eta$ is the local star formation efficiency, and $t_{\rm ff}$ is the local free-fall time. $\eta$ takes into account unresolved physical processes as jets, winds and outflows launched during the process of stellar birth, which limit the gas accretion into stars. Its value is very uncertain, and can span from 0.01 \citep{Krumholz2005a} to 0.3-0.5 \citep{Alves2007, Andre2010}, and in this work we assume $\eta=0.1$. Then it follows that the star formation rate associated to a cell with mass $M_{\rm cell}$ is
\be
\sfr_{\rm cell} = \eta \, \dfrac{M_{\rm cell}}{t_{\rm ff}} \ .
\ee
According to such relation, at every timestep $\Delta t$ of the simulation each cell is assigned a probability of forming a star, given by
\be 
\mathcal{P} = f_\star \, \dfrac{\sfr_{\rm cell} \, \Delta t}{\langle M_\star \rangle} \ ,
\label{prob_sf}
\ee
where $\langle M_\star \rangle \simeq 3.37\,\msun$ is the IMF-averaged mass of the stars and $f_\star$ is a correction factor correcting taking into account that we do not include stars below $1\,\msun$ ($f_\star=N(m>1\,\msun)/N_\textsc{tot}\simeq 0.57$). We also set an $\h2$ density threshold $n_{\rm th}$ for star formation, so only cells with density $n_\h2>n_{\rm th}$ are considered for star formation. The efficiency is then rescaled ($\eta'$) so that the total number of stars formed in a timestep does not depend on the chosen threshold:
\be
\eta' =  \eta \, \left( \sum_i \dfrac{M_{{\rm cell},i}}{t_{{\rm ff},i}} \right) \left( \sum_{n_i > n_{\rm th}} \dfrac{M_{{\rm cell},i}}{t_{{\rm ff},i}} \right)^{-1} \ .
\ee

To ensure mass conservation when a star formation episode occurs, we remove the corresponding star mass from every leaf cell in the cloud (selected as those with $n_\h2 > 10\,\cc$), proportionally to the cell mass:
\be
m'_{\rm cell} = m_{\rm cell} \left(1 - \dfrac{M_\star}{M_{\rm cell,GMC}} \right)
\ee
where $m_{\rm cell}$ and $m'_{\rm cell}$ are the cell gas mass before and after the removal, $M_\star$ is the mass of the newly formed star and $M_{\rm cell,GMC}$ is the total mass of the leaf cells in the cloud. In this way, we remove in total a mass $M_\star$ from the cloud, (1) preferentially removing mass from high-mass, and hence high-density cells, and (2) avoiding the complete depletion of gas mass in cells.

The choice of this star formation routine is alternative to the implementation of sink particles, which form at a minimum mass $m_{\rm seed}$ and then accrete mass from the surroundings. Using sink particles presents additional complications: i) it is not obvious that the correct IMF is recovered, and ii) a reasonable choice for radiation emitted by pre-MS stars have to be made. Since our main goal is to analyse the effect of feedback, and not the star formation process itself, we opted for this simplified recipe.

Finally, we notice that, as we do not track the formation of low-mass stars ($M<1\,\msun$), we do not properly account for the corresponding gas mass depletion. Indeed, such stars are very numerous, and working out the calculation with a Kroupa IMF we obtain that the mass ratio of stars with $M_{<1\msun}$ and $M_{>1\msun}$ is
\[
f_{<1\msun} = \dfrac{M_{<1\msun}}{M_{>1\msun}} \simeq 0.94 
\]
This means that the $\h2$ mass in the simulation is generally overestimated. Therefore, we apply in post-processing the following correction to the $\h2$ mass and the stellar mass to the results of the simulation:
\[
\begin{split}
M'_\h2 &= M_\h2 - f_{<1\msun} M_\star \\
M'_\star &= M_\star + f_{<1\msun} M_\star
\end{split}
\]
where $M_\h2$ and $M_\star$ are the values obtained from the simulation, and $M'_\h2$ and $M'_\star$ are the corrected values.  

\subsection{Radiation from stars}\label{sec:sub:model_radiation}

Newly formed star particles emit radiation as point sources. Every timestep, photons are injected in the cells where stars reside, with an energy spectrum sampled with 10 bins (see Sec. \ref{numerical_simulation}). We neglect the protostellar phase, the pre-main sequence, the red giant branch and following phases, due to their short duration in comparison to the main sequence \citep{kippenhahn1990}. Since both bolometric luminosity and spectrum change with time during the main sequence, for simplicity we take values time-averaged over the lifetime of a star, as detailed below.

We employ the stellar evolutionary tracks for a wide range of star masses from the \parsec~code \citep{Bressan2012}. For each mass in the catalogue, we compute the average bolometric luminosity $L_{\rm bol}$ during the main sequence and then we interpolate to find $L_{\rm bol}$ for the stars in our list. Finally, the stellar spectra are extracted from the Castelli-Kurucz Atlas of stellar atmosphere models \citep{Castelli2003} for the different masses and $L_{\rm bol}$. For reference, on the {\it x} axis of Fig. \ref{star_properties} we report $L_{\rm bol}$ for a range of stellar masses, from 1 $\msun$ to 90 $\msun$. The corresponding spectra are shown in Fig. \ref{star_spectra}, with coloured regions highlighting the $\h2$-dissociating (LW, blue) and $\hi$-ionizing (EUV, red) bands. A star with mass between 3-5 $\msun$ contributes little both to the LW and EUV; a star with mass of 5 $\msun$ starts to contribute non-negligibly to the LW photon budget, with $L_\textsc{lw} \simeq 50 \, \lsun$, and for larger masses ($M\geq 20\,\msun$) the luminosity in the LW saturates to $L_\textsc{lw} \simeq 5\times 10^4 \, \lsun$; finally, only stars with masses larger than 20 $\msun$ have a significant EUV emission, settling to $L_\textsc{euv} \simeq 2 \times 10^5 \, \lsun$ for $M\geq 30\,\msun$.

\subsection{Stellar winds}\label{sec:sub:model_winds}

Besides radiation feedback, massive stars also inject energy in the surrounding ISM  through winds. The mass loss rate ($\dot{M}_{\rm w}$) and the wind kinetic power ($P_{\rm w}$) for the stars in our list are taken from \parsec, averaging over the main sequence. Fig. \ref{star_properties} shows $\dot{M}_{\rm w}$ and $P_{\rm w}$ as a function of mass, for the stars in our sample, color-coded by the time they spend on the main sequence. Stars with mass lower than 10 $\msun$ do not significantly lose mass, so they are not included in the plots. On the other hand, mass loss from very massive stars ($M>50\,\msun$) is about $10^{-5}\,\msun\,{\rm yr}^{-1}$, resulting in a kinetic power of $\sim 10^{37}\,\ergs$ over the their entire lifetime. Assuming an average lifetime of 3 Myr for one of these very massive stars, the total mass loss is $10\,\msun$ with a total energy injection in the ISM around $10^{51}\,\erg$. Hence the energy output is comparable to that of a supernova, and about 40 times the gravitational binding energy of the GMC, showing that stellar winds have potentially enough energy to disrupt the cloud.

Our implementation of stellar winds consists in the injection of mass and energy in the cells adjacent to the star particle, as generally done in grid-based codes \citep{Geen2016, Gatto2017, Haid2019}. 
At each timestep $\Delta t$ of the simulation, and for each star in the box, a mass $\Delta M_{\rm w} = \dot{M}_{\rm w} \Delta t$ is subtracted from the stellar particle and injected in the 27 neighbouring cells (i.e. the $3^3$ cube surrounding the particle host cell) and an energy $\Delta E_{\rm w}=P_{\rm w} \Delta t$ is likewise distributed to those cells. Each cell $j$ receives a different amount of mass ($\Delta M_j$) and energy (either in the form kinetic energy $\Delta E_{{\rm kin},j}$ or thermal energy $\Delta E_{{\rm th},j}$), in order to ensure that the wind is spherically symmetric around the source. For the particle host cell, we inject mass and energy only in the form of thermal energy, since i) we do not resolve the dynamics of the wind inside the cell, and ii) the thermal energy injection is isotropic:
\be
\Delta M_{\rm host} = \dfrac{\Delta M_{\rm w}}{27} 
\qquad 
\Delta E_{\rm th, host} = \dfrac{\Delta E_{\rm w}}{27} 
\ee
Instead for the other 26 neighbouring cells, we inject mass and kinetic energy by
\be
\Delta M = f_j\, \dfrac{\Delta M_{\rm w}}{27} 
\qquad 
\Delta E_{\rm kin} = f_j\, \dfrac{\Delta E_{\rm w}}{27}
\ee
where $f_j$ is the factor accounting for the solid angle covered by the $j$-th cell when seen from the source. If we consider that all the neighbouring cells are always kept at the maximum refinement level, and we enforce
\be
\sum_{j=1}^{26} f_j = \dfrac{26}{27},
\ee
we obtain 
\be
f_j = \dfrac{39}{22} \, \left( \dfrac{\Delta x}{r_j} \right)^2
\ee
where $\Delta x$ is the cell size and $r_j$ the distance from the host cell. 
These cells receive a kick in velocity in the direction of the line joining the centres of the $j$-th cell and of the host cell.

\subsection{Supernovae}\label{sec:sub:model_SN}

Stars with mass larger than 8 $\msun$ explode as Type II \citep{Smartt2009} supernovae at the end of their life. For simplicity, we neglect all the stellar evolution phases following the main sequence (as the red giant branch, the horizontal branch and the asymptotic giant branch), that are expected to represent only a short fraction of the total stellar lifetime, and we assume that stars explode just after the main sequence. The main sequence time is taken from the evolutionary tracks in \parsec. When a star with mass $M$ explodes as a supernova, it injects in the surrounding cells a mass $M_{\rm ej}=M-M_{\rm rem}$, where $M_{\rm rem}=1.4\,\msun$ is the mass of the remnant \citep[the Chandrasekhar limit,][]{Chandrasekhar1939}. After the explosion, we set the particle mass to $M_{\rm rem}$ and assume no further radiation is released. The ejecta are distributed among the 27 neighbouring cell in the same fashion of the winds, but with a total energy input of $10^{51}$ erg: $1/27$ of the energy is injected as thermal energy in the host cell of the star, and the remainder $26/27$ as kinetic energy in the surrounding cells. 


\section{Results}
\label{results}

\begin{figure*}
\centering
\includegraphics[width=0.95\textwidth]{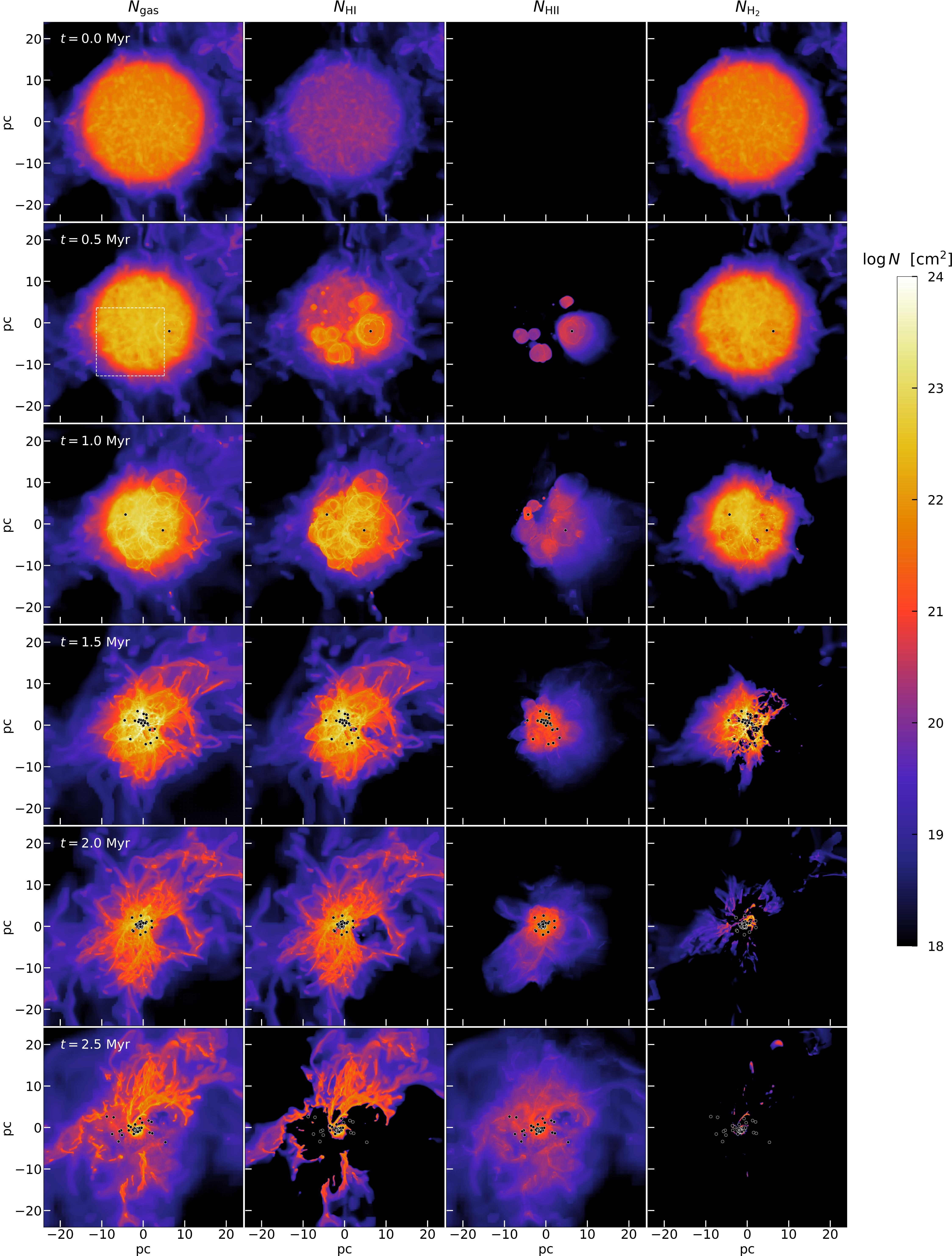}
\caption{Evolution of the molecular cloud since the epoch $t_0$ of the formation of the first star. The four columns show respectively the total, $\hi$, $\hii$ and $\h2$ column densities at different times ($t=0,\,0.5,\,1.0,\,1.5,\,2.0,\,2.5$ Myr). The black dots represent the locations of stars more massive than $30\,\msun$. As stars form, molecular gas is converted in atomic form, and ionized in proximity of massive stars, with the result that the cloud is destroyed around 2.5-3 Myr. The dashed white square in the second left panel is the zoomed region in Fig. \ref{hiiregion_slices}.
\label{gmc_panel}}
\end{figure*}

\begin{figure*}
\centering
\includegraphics[width=\textwidth]{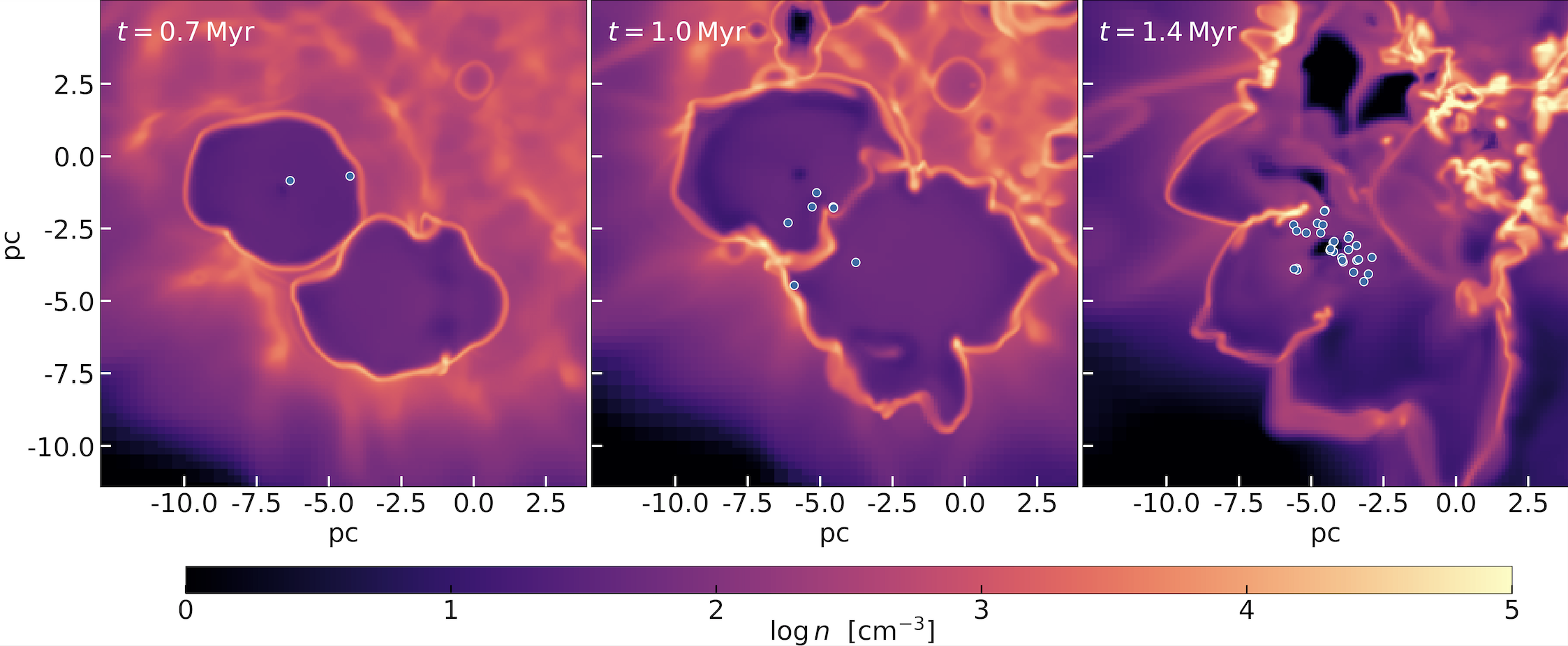}
\caption{Slices of the cloud at different times ($t=0.7$ Myr, 1.0 Myr and 1.4 Myr), zoomed to show the evolution of two HII regions (the zoomed area is marked with a white dashed square in Fig. \ref{gmc_panel}). The maps are color-coded according to the gas number density, and the dots show stars of any mass within 0.1 pc from the slice. From left to right, we can see the formation of two HII regions, the development of instabilities at the edges, and the formation of dense clumps with a prolonged tail.
\label{hiiregion_slices}}
\end{figure*}

\subsection{Overview of cloud evolution}
\label{subsec:overview}

Images of the cloud are represented in Fig. \ref{gmc_panel}, showing snapshots taken at regular intervals up to $t=2.5$ Myr (we recall that time is counted starting from $t_0= 3$ Myr, the time at which star formation is enabled). The four columns, from left to right, show the total gas surface density and the surface density of $\h2$, $\hi$ and $\hii$.

At $t=0$ Myr, the cloud has developed a complex structure of filaments and clumps, forming where the initial turbulent field produces an enhancement in the gas density. Then, stars form stochastically in the overdense cells, according to the probabilistic recipe described in Sec. \ref{star_formation}. Low-mass stars (1-2 $\msun$) are able to form small partially ionized gas bubble around them, with a thick transition region to neutral and molecular hydrogen. Most massive stars ($>30\,\msun$) form large HII regions which are clearly visible in the surface density plots. In particular, notice the main bubble found in the $t=0.5$ Myr snapshot, centred in (6.5 pc, -2 pc), due to the formation of a 40 $\msun$ star. The area inside the bubble is quickly emptied, due to the combined effect of radiation and winds.

As the cloud evolves ($t\geq 1.5$ Myr), the high density gas which has managed to resist feedback keeps collapsing towards the centre of the box, while less dense filaments protrude at larger distances. Since SFR is proportional to gas density, it increases to very high values during the final stages of the simulations ($\sim$ 0.1 $\msun\,\yr^{-1}$), consuming very rapidly the gas content of the most dense clumps. Hence, feedback and star formation are the two mechanisms decreasing the amount of molecular gas, consuming it all by $t\simeq2.8$ Myr. Since even the most massive stars in the simulations have a lifetime of about $\sim 3$ Myr, the cloud has already ceased to exist before any supernovae get the chance of exploding. 

While from $t=1.5$ Myr the molecular gas is only concentrated in clumps in the central region of the cloud, HI maps show that the atomic hydrogen is not collapsing towards the centre. Maps at $t=1.5$ Myr and $t=2.0$ Myr show indeed the an HI cloud with radius $>$ 10 pc persists, as the gravitational collapse prevented by the high temperature of FUV-heated gas ($T\simeq 10^3$ K). However, in the last snapshot ($t=2.5$ Myr), the dense molecular clumps are completely dissociated and ionizing radiation is free to propagate in the whole cloud. As a result, the HI gas is almost completely ionized at the end of the simulation.

\subsection{Structure of HII regions}
\label{subsec:hii_regions}

In order to study the structure of HII regions, we show in Fig. \ref{hiiregion_slices} density slices centreed around two close-by massive stars (25 $\msun$ and 40 $\msun$) forming at $t=0.02$ Myr and $t=0.32$ Myr. All the stars within 1 pc from the cut are shown as black dots. The slices are taken at different times, to show the evolution of the dense shell of gas developing at the edge of HII regions and the formation of clumps with size $<$ 0.1 pc. 

In the leftmost slice ($t=0.7$ Myr), the HII region are already well developed. Fully ionized gas in the bubbles has a density around $n\sim 100\,\cc$, while gas at the edge reaches around 2000 $\cc$, due to the piling-up of gas pushed by the radiation-induced shocks and the winds. This shell is HI-rich, and represents indeed a PDR, formed thanks to FUV radiation getting through the ionized region and dissociating $\h2$. From the slice, it is evident that the dense shells are not regular in shape and density, due to inhomogeneities in the gas prior to the ionization front propagation. 

In the second slice ($t=1.0$ Myr) the shells fragment due to instabilities, triggered also by the collision between the two close-by shells, inducing the formation of small and dense clumps characterized by a dense and compact head followed by a tail. These clumps move towards the centre of the box due to gravitational potential of the cloud (rightmost slice, $t=1.4$ Myr). Due to their high density, stellar feedback is ineffective in destroying such clumps, resulting in their rapid conversion into new stars, a process that slowly consumes all the available molecular gas.

Another noticeable effect seen in the simulation is the formation of stars in the compressed shell around $\hii$ regions, as seen in both in the first and in the second slice. Since the probability of forming star is proportional to gas density, the formation of dense shells around $\hii$ regions fosters the birth of new stars. This scenario of self-propagating (or triggered) star formation has been frequently investigated in previous works, both from an observational point of view \citep{Lada1999, Zavagno2006, Deharveng2006, Deharveng2010, Beerer2010, Liu2016, Yep2020} and from simulations \citep{Hosokawa2005, Hosokawa2006, Hosokawa2006a, Dale2007a, Dale2007b, Walch2013}. 

\subsection{ISM phases evolution}\label{subsec:ism_phases}

\begin{figure*}
\centering
\includegraphics[width=\textwidth]{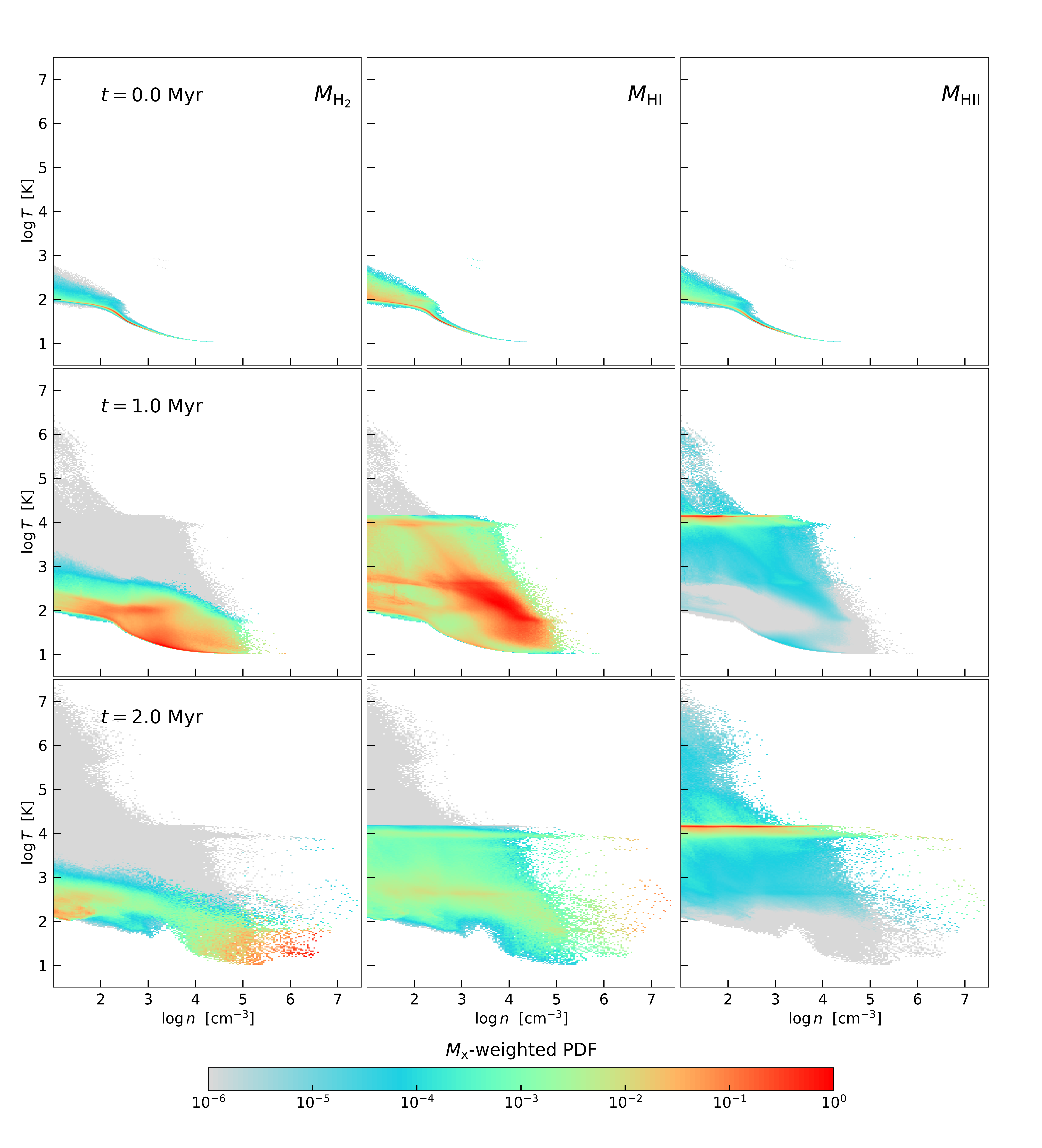}
\caption{Phase diagrams of the gas in the cloud. The three rows show different evolutionary stages, at $t=0$, $1$ and $2$ Myr respectively. Each panel shows a $M_{\rm x}$-weighted PDF in the density-temperature ($n-T$) plane, where $M_{\rm x}$ is the mass of the species X. The three columns correspond to the species $\h2$, $\hi$ and $\hii$. As the cloud evolves with time, molecular gas is converted into atomic and ionized gas.}
\label{eos_panel}
\end{figure*}

\begin{figure}
\centering
\includegraphics[width=0.48\textwidth]{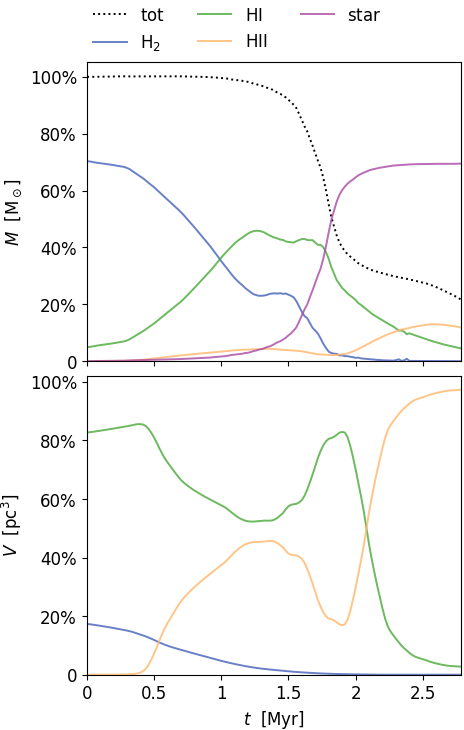}
\caption{{\bf Upper}: Time evolution of $\h2$ mass, $\hi$ mass, $\hii$ mass, stellar mass and total gas mass (including also helium) in the whole computational box (hence including both the cloud and the surrounding low density medium). {\bf Lower}: Time evolution of the volume occupied by $\h2$, $\hi$ and $\hii$ in whole box; a cell is included in one phase if more than 50\% of its mass is in such a phase. 
\label{mass_evolution}}
\end{figure}

\begin{figure*}
  \begin{subfigure}[b]{0.49\textwidth}
    \includegraphics[width=\textwidth]{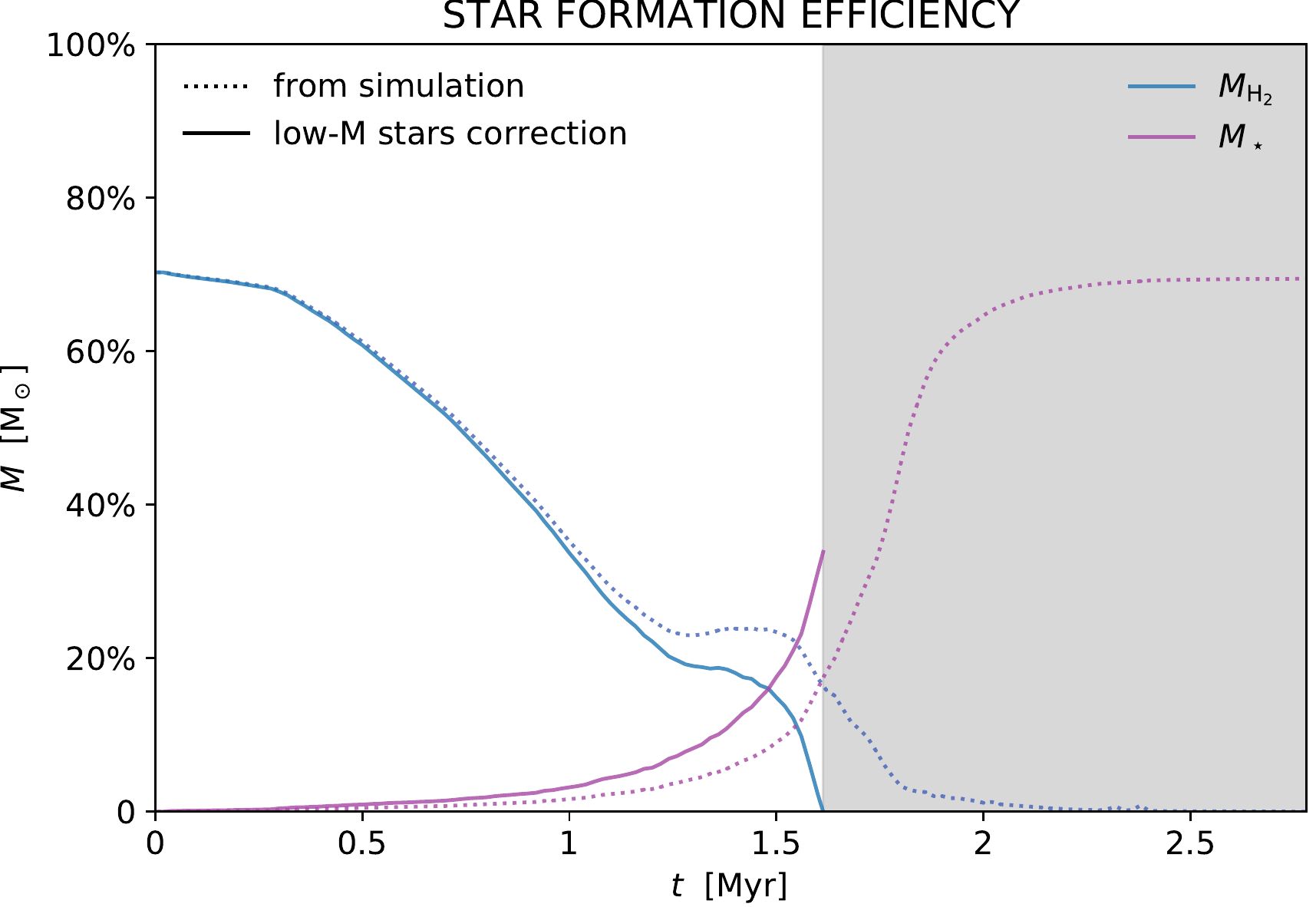}
  \end{subfigure}
  \begin{subfigure}[b]{0.49\textwidth}
    \includegraphics[width=\textwidth]{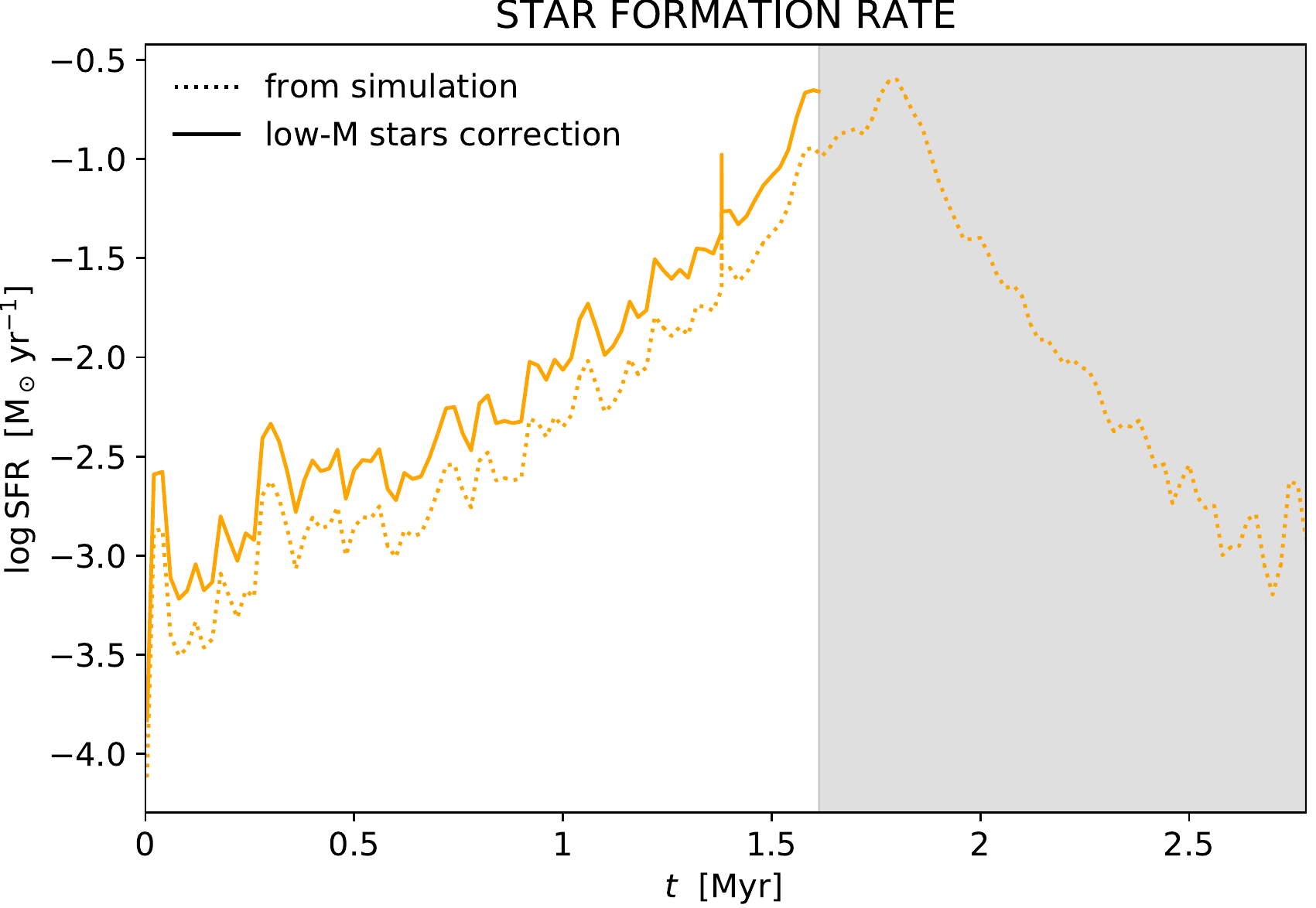}
  \end{subfigure}
  \caption{{\bf Left}: time evolution of $\h2$ mass and stellar mass, and relative correction accounting for low mass stars ($M<1\,\msun$), which are not included in the simulation. Accounting for this correction, the cloud is destroyed in a shorter time ($t\simeq 1.6$ Myr), and the final SFE is lower ($\sim 36 \%$). {\bf Right}: time evolution of the star formation rate, again plotted together with the low mass star correction. The shaded area mark the times at which the cloud is destroyed (i.e. there is no $\h2$ left) according to the low-M star correction.
  \label{sfe_sfr}}
\end{figure*}

The evolution of different gas phases in the whole computational box can be studied with density-temperature phase diagrams, shown in Fig. \ref{eos_panel}. Different rows correspond to simulation snapshots at $t=0$ Myr, $t=1$ Myr and $t=2$ Myr. Colors represent the normalized PDF, weighted by the mass of $\h2$, $\hi$, and $\hii$, respectively. At $t=0$ Myr, the gas in the box is $\simeq70\%$ molecular, with the rest being either in atomic ($\sim 5 \%$) or ionizd form. This happens because gas compression (shocks and gravitational collapse) increases the temperature, fostering collisional dissociation of $\h2$ and ionization of $\hi$. On the other side, the gas filling the box outside the cloud has $n<10^2\,\cc$ and $T\simeq 10^{3-4}$ K, either in atomic or ionized form.

The row at $t=1$ Myr shows features due to the presence of stars, heating the gas via both radiation and winds. At this stage, most molecular gas ($\sim75\%$ of the total molecular gas) is present in the form of cold ($T<30$ K) and dense ($n>10^{3-4}\,\cc$) clumps, even if there is some less dense molecular gas at $n<10^3\,\cc$. The $M_\hi$-weighted map is populated mainly by gas in the range $n=10^3-10^4\,\cc$, with temperatures varying from $10^2$ K to $10^4$ K. Two horizontal branches ($T\simeq 10^3$ K and $T\simeq 10^4$ K) are visible, and correspond to the edges of HII regions, where the gas is compressed by radiation-driven shocks and EUV photons are not able to penetrate. Warm medium with $T < 500$ K is associated with PDRs, and contains around $20\%$ of HI. The $M_\hii$-weighted map shows a clear horizontal line at $T\simeq 4\times 10^4$ K, due to photoionization heating of hydrogen and helium. These ionized regions have densities ranging from $10\,\cc$ to $10^3\,\cc$. Gas is heated to even higher temperatures ($T\simeq 10^{5-6}$ K) around massive stars, due to the kinetic energy injection by winds.

The last row ($t=2$ Myr) shows an advanced evolutionary stage of the cloud, where most of molecular gas has been dissociated. $\h2$ survives only in very dense clumps ($n>10^4\,\cc$) and a few patches of low density gas, which are shadowed from radiation. Also, $\hi$ is much less abundant than in the previous snapshot, due to photoionization. Indeed, the horizontal branch in the $M_\hii$-weighted map is much more extended in this case (up to $10^4\,\cc$).

\subsection{SFE and SFR}\label{subsec:sfe_and_sfr}

The upper panel of Fig. \ref{mass_evolution} shows the time evolution of gas mass in the box in the different phases ($M_\h2$, $M_\hi$ and $M_\hii$), together with the total gas mass $M_{\rm tot}$ (including also all the other species, as ${\rm He}$) and the stellar mass $M_\star$. All masses include all the gas in box, so both the GMC and the surrounding gas.

As the cloud evolves, molecular gas has two possible fates: (1) it is converted into atomic and ionized gas by stellar feedback, (2) it goes into stars. This means that the efficiency of stellar feedback determines the molecular gas mass which remains available for star formation. $M_\h2$ goes to zero around $t_{\rm ev} \simeq 2.2$ Myr, marking the complete evaporation of the GMC. $\hi$ mass increases as $\h2$ is dissociated, peaking between $1.2$ and $1.7$ Myr. Later, $\hi$ is ionized and hence its mass tends to zero. $\hii$ mass grows as more ionizing flux is produced, and at the end of the simulation the mass is either in in stars or $\hii$, with a small percentage in $\hi$.

The volume occupied by the different gas phases is shown in the lower panel of Fig.~\ref{mass_evolution}. The volume occupied by the species X (HI, HII or $\h2$) is defined as the sum of the volume of all cells where the mass fraction $\mu_{\rm x} = M_\textsc{x} / (M_\hi + M_\hii + M_\h2)$ is larger than 50\% 
\footnote{We have verified that changing the threshold to 75\% or 90\% does not yield any significant difference in the trend. We also obtain a similar result by removing the threshold and weighting the volume of the cells by $\mu_{\rm x}$.}. 
At $t=0$ Myr, the gas in the cloud is in fully molecular form ($\sim$ 20\% of the total volume, while the surrounding ISM is in atomic phase. Then, the volume of the molecular phase decreases with time, both because of dissociation by LW photons and gravitational collapse. In the meantime, the external medium is ionized, explaining the decrease in HI volume and the increase in HII volume between $t=1$ Myr and $t=1.5$ Myr. 
The cloud collapse proceeds in the interval $t\simeq 1.5-2$ Myr, corresponding to the cloud free-fall time, with the formation of more dense clumps. As a result, dense gas absorbs ionizing photons and the size oh HII regions reduces, and this can be seen in Fig.~\ref{mass_evolution} as a dip in HII volume and a correspondent peak in HI volume.
After 2 Myr, dense clumps are consumed by star formation, and ionizing radiation can again propagate in the whole computational box, filling almost all the volume. 

When the cloud is completely evaporated, the total stellar mass is $M_\star\simeq 7.4\times 10^4 \msun$, which means a global ${\rm SFE}= M_{\star}/M_{\rm GMC}\simeq 74\%$. Nevertheless, the correction for the fact that we neglected the formation of low-mass stars (and hence the corresponding mass removal) must be applied, as detailed at the end of Sec. \ref{star_formation}. The left panel of Fig. \ref{sfe_sfr} shows with dotted lines the values of $\h2$ mass and stellar mass straight from the simulation, and the corrected values with solid lines. We can see that the corrected value $M'_\h2$ approaches zero around $t=1.6$ Myr, and by this time the total stellar mass is $M'_\star\simeq 3.6\times M_{\star}$, or SFE$\simeq 36\%$.

\subsection{Radiation in the GMC}\label{sec:radiation_field_analysis}
\label{subsec:radiation}

\begin{figure*}
\centering
\includegraphics[width=\textwidth]{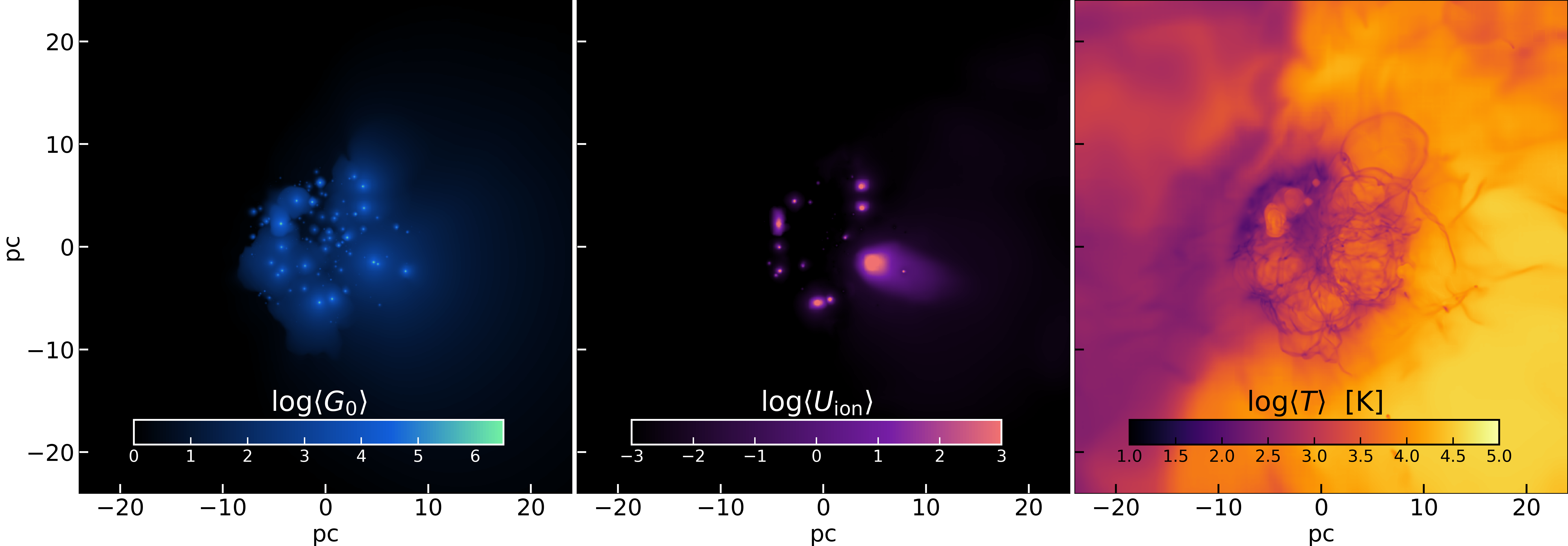}
\caption{Snapshots of the cloud properties at $t=1.0$ Myr: from left to right, projections of (a) FUV flux $G_0$ weighted by photon number, (b) ionization parameter $U_{\rm ion}$ weighted by photon number and (c) temperature $T$ weighted by gas mass. At this stage, photons have not leaked yet from the cloud, and the flux is high only close to massive stars or stellar clusters.
\label{gmc_radiation}}
\end{figure*}

\begin{figure*}
\centering
\includegraphics[width=0.49\textwidth]{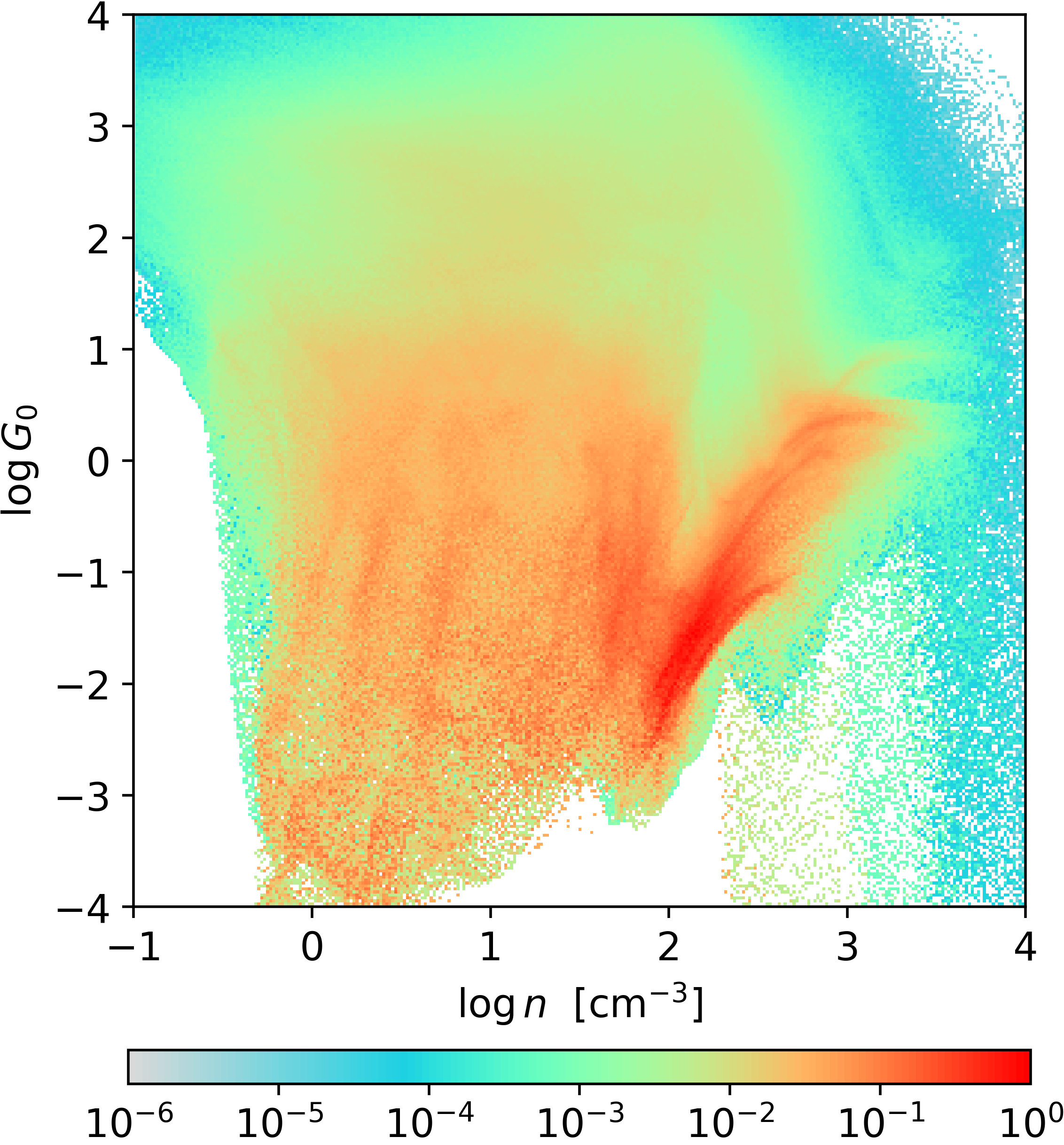}
\includegraphics[width=0.49\textwidth]{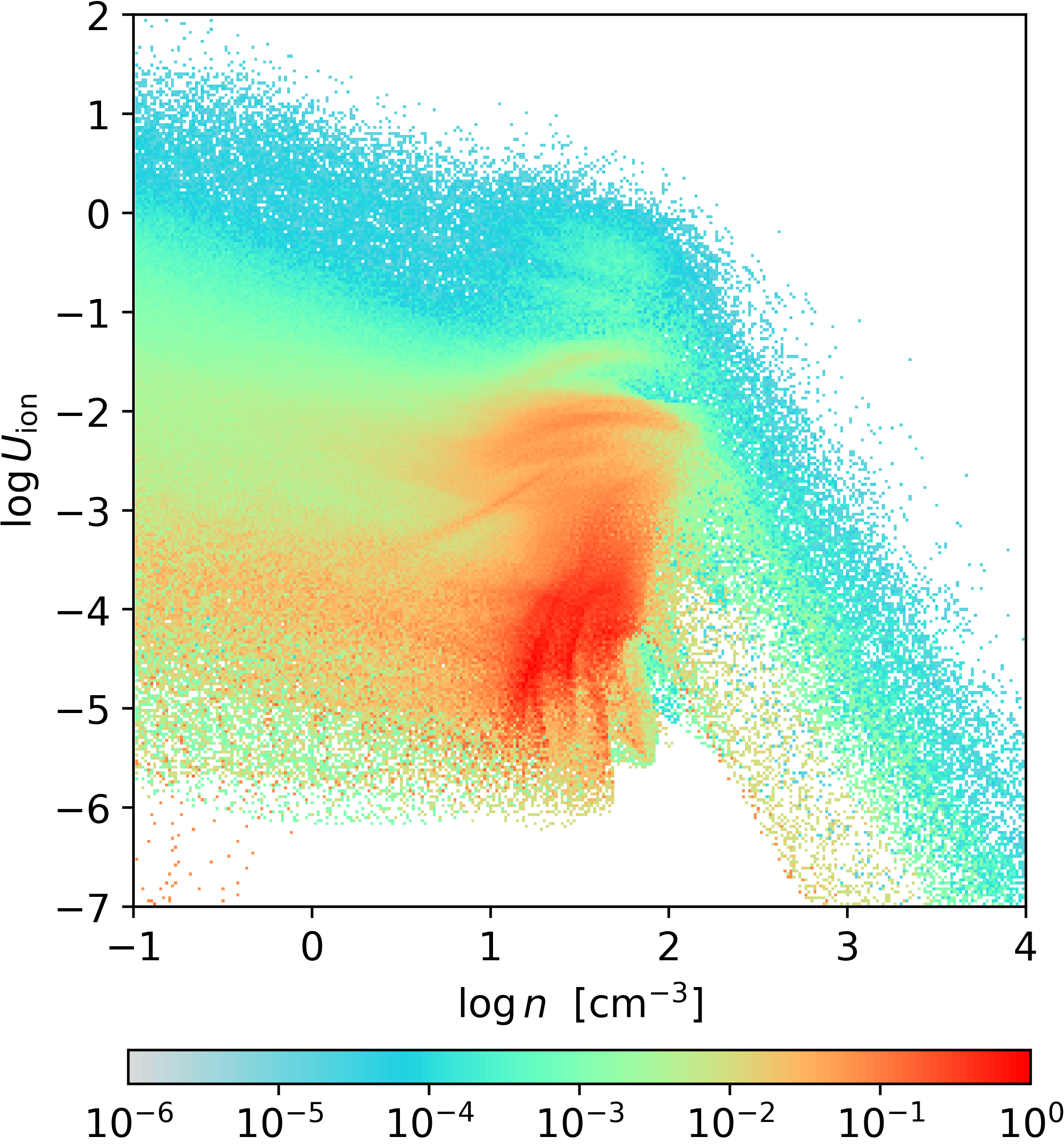}
\caption{
$G_0-n$ (left) and $U_{\rm ion}-n$ (right) phase-diagrams at $t=1$ Myr. The PDF are volume-weighted and then normalized. Regions with the density ($n>10^3$) may have a high FUV flux ($G_0>10^2$), while they are completely self-shielded from ionizing radiation. At low gas densities ($n<10^2$), $G_0$ can assume a variety of values from $10^{-3}$ to $10^3$, while the ionization parameter is generally $>10^{-3}$. 
\label{rad_phase_diagram}}
\end{figure*}

\begin{figure}
\centering
\includegraphics[width=0.49\textwidth]{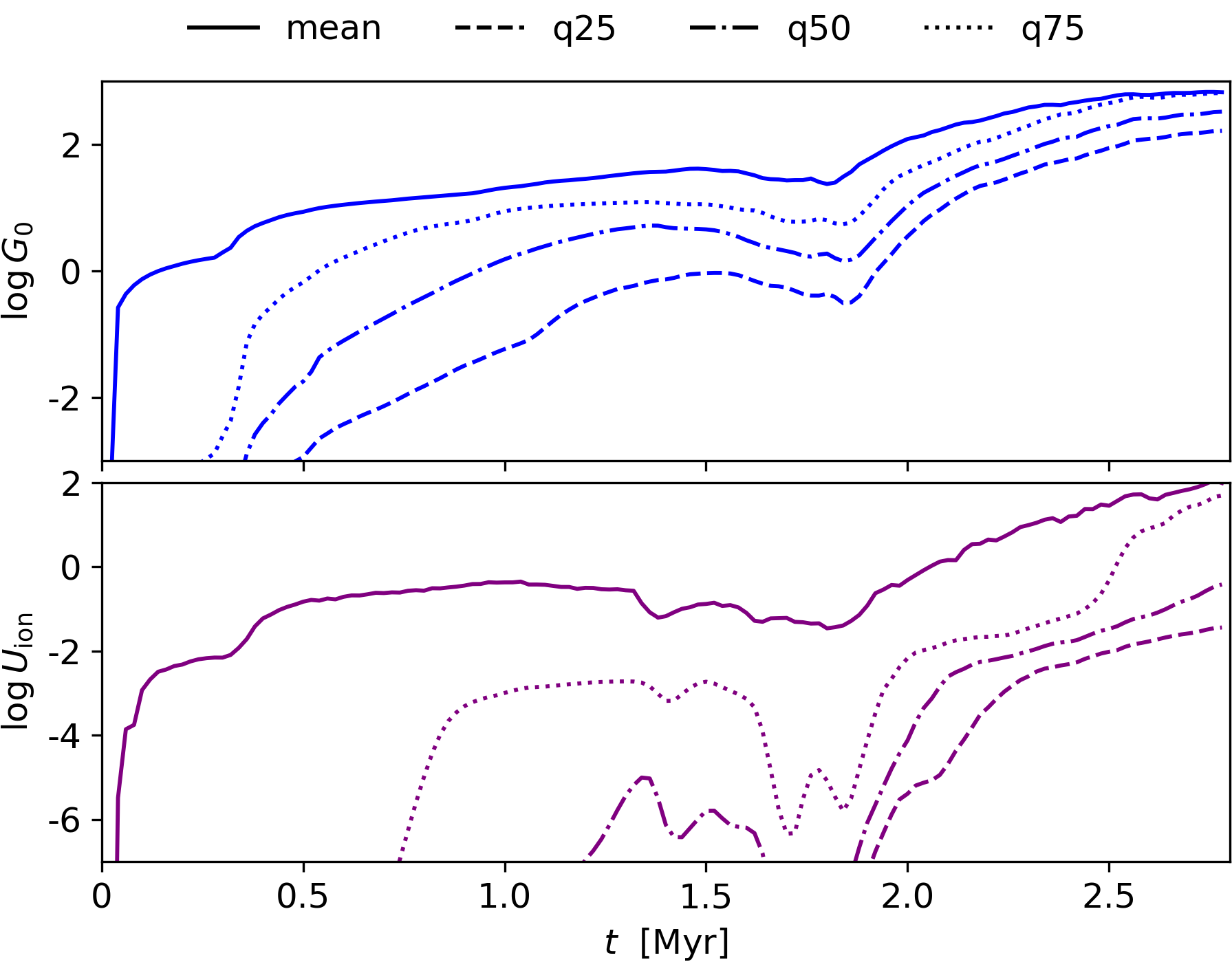}
\caption{ 
Time evolution of $G_0$ ({\bf upper} panel) and $U_{\rm ion}$ ({\bf lower} panel) in the cloud. The solid line shows the evolution of the volume-weighted average value, while the dashed, dotted-dashed and dotted lines show the 25th, 50th and 75th volume-weighted percentiles respectively ($q_{25}$, $q_{50}$ and $q_{75}$). The larger scatter of the $U_{\rm ion}$ distribution indicates that the ionization field is more patchy than the Habing one. 
\label{rad_evolution}}
\end{figure}

\begin{figure}
\centering
\includegraphics[width=0.49\textwidth]{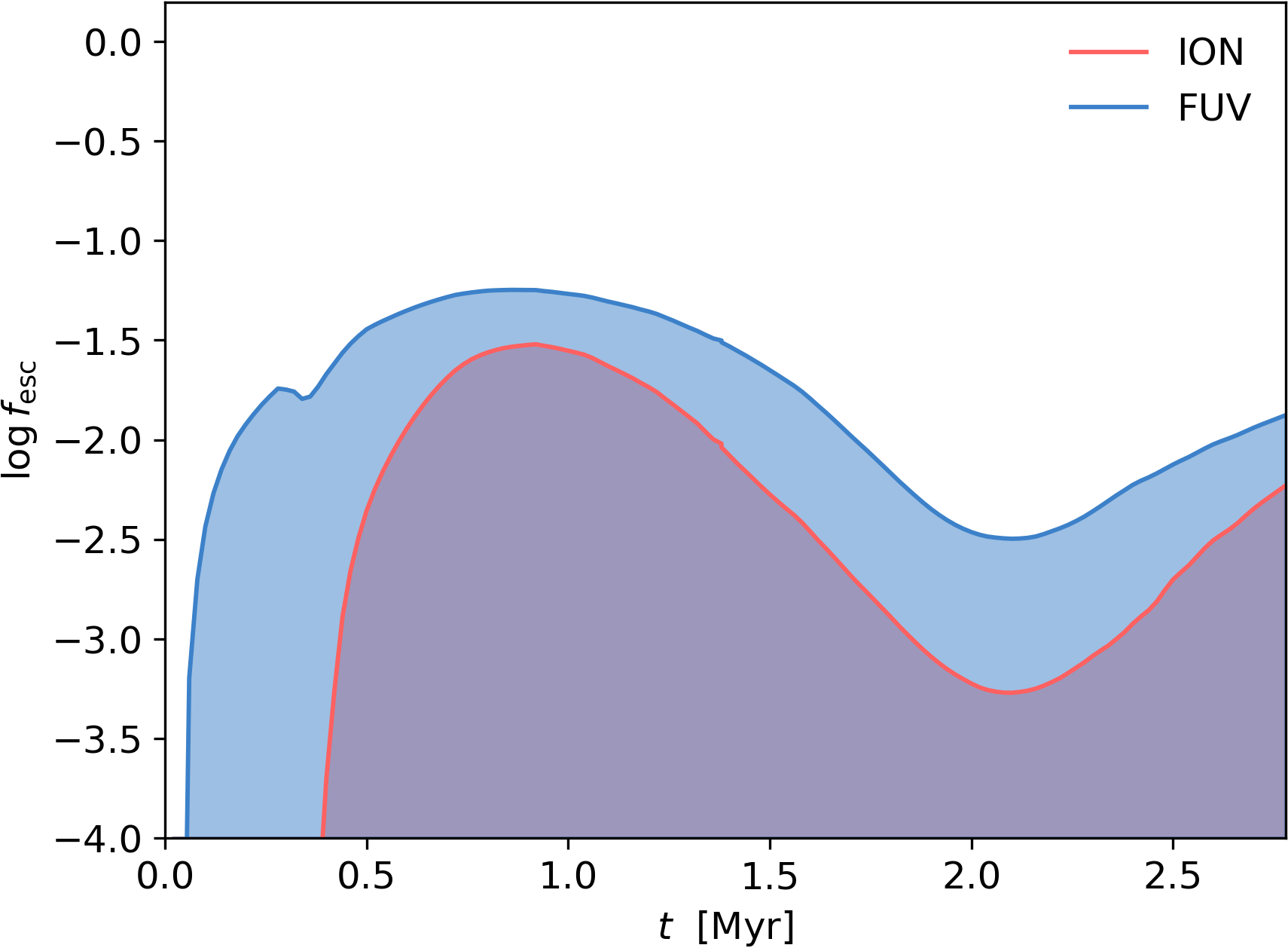}
\caption{ 
Time evolution of the escape fraction (see definition in eq. \ref{fesc}) of FUV (blue line) and ionizing (red line) photons.  
\label{f_escape}}
\end{figure}

Fig. \ref{gmc_radiation} shows the maps of photon density-weighted Habing flux $G_0$ (6 eV $<h\nu<$ 13.6 eV), photon density-weighted ionization parameter ($U_{\rm ion}$, defined as the ratio between ionizing, $h\nu>$ 13.6 eV, photon density and gas density) and gas mass-weighted temperature $T$ at $t=1.0$ Myr. The flux is high in the neighbourhood of massive stars, but it has not leaked yet from the whole cloud. Typical values of $G_0$ are $10^{4-5}$ and $U_{\rm ion}\simeq 10^3$ in the immediate proximity of massive stars, as they have devoided of gas their surrounding.
The map shows mass-weighted temperatures of 3000~K inside the cloud, however the maximum temperatures inside HII regions can reach values of $10^{7-8}$~K near very massive stars, because of the mechanical feedback.

An analysis of the radiation field in molecular clouds is particularly relevant for cosmological simulations of galaxy formation, where generally it is possible only to marginally resolve the internal structure of molecular clouds \citep{Leung2019}. With a typical resolution of tens of parsecs \citep{Rosdahl2018,Pallottini2019}, a cell represents a whole GMC and therefore its properties are averaged, so local inhomogeneities cannot be taken into account. GMC emission properties -- in particular far-infrared lines, as [CII], [NII], and [OIII] and CO -- has been shown to be sensitive to their internal density structure \citep{Vallini2017,Vallini2018}, while the radiation field is typically assumed to be constant. In particular, [CII] emission depends on $G_0$, while [NII] and [OIII] are strongly affected by  $U_{\rm ion}$. Hence, the approximation of an average uniform $U_{\rm ion}$ (due to resolution limits) can lead to an incorrect estimation of line emission (see discussion in \citealt[][]{Olsen2018} and \citealt{Pallottini2019}).

In Fig. \ref{rad_phase_diagram} we plot the $G_0-n$ and $U_{\rm ion}-n$ phase diagrams at $t=1$ Myr, showing the normalized volume-weighted PDF. High density gas ($n>10^3\,\cc$) has $U_{\rm ion}\simeq 0$, being able to self-shield from ionizing radiation, while $G_0$ can attain large values $G_0>10$. On the other hand, low density gas ($n<10^2\,\cc$) typically shows $G_0>10-100$ and ionization parameter $U_{\rm ion}\simeq 10^{-2}$, as this gas is associated with $\hii$ regions. According to \citet{Pallottini2019} [CII] emission mostly comes from regions in which $n\simeq2\times 10^2\,\cc$, $G_0\simeq 20$, and $U_{\rm ion}$ is relatively low, implying that they are neutral \citep[see also][]{Ferrara2019}. Lines as [NII] and [OIII] are instead emitted from ionized regions with densities around 50-100 $\cc$ and $U_{\rm ion}>10^{-3}$. 

The time evolution of $G_0$ and $U_{\rm ion}$ in the cloud are shown in Fig. \ref{rad_evolution}: the different lines show the volume-weighted average and the 25th, 50th, and 75th volume-weighted percentiles\footnote{The 25th volume-weighted percentile is defined as the value $q_{25}$ such that $q<q_{25}$ in 25\% of volume. Analogously for the 50th percentile, i.e. the median, and the 75th percentile.}. The average values of both $G_0$ and $U_{\rm ion}$ increase almost steadily during the simulation, showing a slight drop around $t\simeq 1.5 - 2$ Myr, corresponding to the increase in $\hi$ seen in Fig. \ref{mass_evolution}. At the end of the simulation, the volume-weighted $\langle G_0 \rangle$ stabilizes at a value of $\sim 10^3$, and $\langle U_{\rm ion} \rangle$ reaches $10^2$.
The spread in the percentiles evolution is narrower for $G_0$ than $U_{\rm ion}$: defining the relative spread between percentiles as $\delta X = (q_{75} - q_{25}) / q_{50}$, we have $\delta U_{\rm ion}$ larger than $\delta G_0$ by a factor $10^7$ at $t=1$ Myr and by a factor $\sim 100$ for $t>2$ Myr. This means that the distribution of $G_0$ at a given time is more peaked, while the distribution for $U_{\rm ion}$ is more spread with a larger variance. Hence, prediction on the intensity of [NII] and [OIII] lines will be strongly affected by the actual distribution of $U_{\rm ion}$, and the assumption of a uniform ionizing field appears not to be satisfied in our cloud. A more detailed analysis of emission properties from our simulated GMC is postponed to a subsequent work.  

Finally, we briefly discuss the time evolution of the escape fraction. This is defined in band X, at time $t$, as the ratio 
\be
f_{\rm X,esc}(t)=\dfrac{N_{\rm X,esc}(<t)}{N_{\rm X,prod}(<t)}
\label{fesc}
\ee
where:
\begin{itemize}
\item $N_{\rm X,esc}(<t)$ is the total number of photons in band X which have crossed the spherical surface of radius $R_{\rm s}=20$ pc (i.e. the initial cloud radius) up to time $t$;
\item $N_{\rm X,prod}(<t)$ is the total number of photons in band X produced by all stars in the cloud up to time $t$.
\end{itemize}
In Fig. \ref{f_escape} we show the escape fractions in the FUV band (blue line) and in the ionizing band (red line). FUV photons manage to escape earlier from the cloud, with $f_{\rm FUV,esc} > 0.01$ already around $t\simeq 0.2$ Myr, since they are able to propagate to larger distances in molecular gas before being absorbed. For the same reason, $f_{\rm FUV,esc}(t)> f_{\rm ion,esc}(t)$ at any time. The drop in $f_{\rm esc}$ in both bands around 2 Myr corresponds to the increase in mass and volume of the neutral phase, apparent from Fig. \ref{mass_evolution}. Later on, $f_{\rm esc}$ increases again, and we expect it to increase monotonically to large values while neutral gas is progressively ionized. Note, however, that the ionizing photon production goes to zero when massive stars (the main contributors to $N_{\rm ion,prod}$) explode as SNe ($t\apprge 3$ Myr). We find that throughout the evolution  $f_{\rm ion, esc} \simlt 0.03$, also in agreement with similar studies \citep{Howard2018a, Kimm2019, He2020}.


\section{Discussion}\label{sec:discussion}

Our simulation shows how stellar feedback shapes the structure of a GMC, by ionizing, heating and pushing the gas. HII regions form as bubbles around massive stars, finally merging and leaving molecular gas concentrated in dense clumps with tails (see 5th row in Fig. \ref{gmc_panel}). Similar substructures are found in other works featuring photoionization feedback \citep[e.g][]{Walch2015, Haid2019, He2019}. Nevertheless, the quantitative results on the initial mass function (for simulations featuring sink particle formation), cloud dispersal time and star formation efficiency are sensitive on differences on the initial setup. For example, \citet{Geen2018} show that simulations with different initial random seed of the turbulent velocity field achieve different values of the SFE (from $\sim6\%$ to $\sim 21\%$). Similarly, \citet{Haid2019} evolve two clouds with comparable mass ($\sim 10^5 \,\msun$) by zooming-in different regions of a galaxy-scale simulation: the clouds show different morphology (only one cloud shows a very dense blob at the center), SFEs (6\% and 13\%) and dispersal time (only one cloud is dispersed in the simulation time, 3 Myr).

Other recent works on GMC simulations \citep[e.g][]{Raskutti2016, Howard2017, Kim2018} find SFE$\simeq 10-20\%$ for clouds with similar properties to this work ($M=10^5\,\msun$, $\bar{n}=120\,\cc$), despite not including all feedback mechanisms simultaneously (photoionization/dissociation, radiation pressure, winds). This is mainly due to a few significant differences with respect to our setup, which result in more spread and less dense clouds: (1) most works relax the cloud in a non-turbulent surrounding medium, (2) gravity is turned off or a reduced gravitational constant is used in the relaxation stage, (3) different initial cloud profiles are adopted (isothermal profiles, Bonnor-Ebert spheres, etc.). As a consequence, these clouds are less star forming and more subject to gas evaporation by stellar feedback. Instead, in our simulation, we start with a uniform, self-gravitating cloud, confined by a turbulent surrounding medium. Star formation is enabled after 3 Myr, when the cloud gas is denser and has already started collapsing towards the centre of the box. Hence, it is natural to expect a more rapid star formation and a higher SFE in our case. In addition, stellar feedback is sometimes implemented in a different way. For instance, \citet{He2019} assume a constant light-to-mass ratio for sink particles, hence overestimating the UV photons emitted by low mass stars (as shown in Fig. \ref{star_spectra}, only stars with mass $M>10\,\msun$ have a relevant contribution both in the FUV and the EUV).

Nevertheless, it is interesting to notice that both in this and previous theoretical works, the SFE is typically larger than what typically observed \citep[$\sim 1-10\%$][]{Lada2010, Murray2010, Lee2016, Lada2016, Ochsendorf2017} in local GMCs in the Milky Way, despite the accurate implementation of stellar feedback. A possible explanation is the absence of turbulence driving in our simulation, meant to mimic external shear and compression due to the environment in which the cloud could be embedded in (turbulent motion in the spiral arms of the galaxy, supernova explosions in the neighbourhood of the GMC, etc.). 

In addition, the inclusion of magnetic fields could also decrease the SFE of the cloud, since magnetic pressure entails lower clump densities. As a result, clumps are less star forming and more easily dissolved by radiation. \citet{Federrath2013} have analysed the impact of different effects on star formation, finding that the effect of an increase of the turbulence is twofold: (1) the increase of $\alpha_{\rm vir}$ stabilizes the cloud against collapse, decreasing the SFE, (2) compressive modes with higher Mach number $\mathcal{M}$ foster star formation by creating local compression; moreover, the presence of solenoidal modes (keeping $\mathcal{M}$ constant) decreases the SFE at least of an order of magnitude, while magnetic fields with Alfvenic turbulence with $\mathcal{M}_{\rm A} \sim 1.3$ decrease the SFE by a factor of two. Hence, the inclusion of turbulence driving and magnetic fields in our simulation would help to obtain lower values for the SFE, that are closer to the observed ones. We plan to address this issue in a future work.


\section{Conclusions}\label{sec:conclusions}

We have studied the evolution of a typical Giant Molecular Cloud (GMC) by running a 3D radiative transfer (RT), zoom hydro-simulation, including a full chemical network tracking H$_2$ formation and dissociation, and following individual stars forming inside the cloud. Multiple feedback mechanisms, such as photodissociation/ionization, radiation pressure, stellar winds and supernovae, are included simultaneously.

The simulation has been run with the RT version of the \ramses~code \citep{Teyssier2002, Rosdahl2013}, by coupling the RT module with the non-equilibrium chemical network generated with \krome~\citep{Grassi2014} in the same fashion as \citet{Decataldo2019} and \citet{Pallottini2019}. We have implemented a stochastic star formation recipe, drawing stars from a \citet{Kroupa2001} Initial Mass Function (IMF) and placing them in the simulation already in main sequence. To save computational time, we consider only stars with masses larger than $1\,\msun$ and then apply a correction to account for the mass consumption by low-mass star formation.
Each star emits radiation with a spectrum depending on its mass, sampled with 10 photon bins, and injects momentum isotropically in the interstellar medium via winds. The simulated GMC has a mass $M=10^5\,\msun$, radius $R=20$ pc and initial virial parameter $\alpha_{\rm vir}=2$, meaning that the cloud is initially unbound.

The cloud presents dense clumps and filaments before star formation, due to turbulence and gravitational instabilities. When stars form, they affect the surrounding environment. Thanks to the inclusion of different mechanisms, and the accurate treatment of radiation and chemistry, it has been possible to identify different phases in the cloud evolution. In a first phase (up to $t=1$ Myr), we assess the effect of radiation and winds on the structure and composition of the cloud. Ionized bubbles develops around massive stars (with densities $n\simeq 100\,\cc$ and $T\simeq 4\times 10^4$ K) with dense edges ($n\simeq 10^3\,\cc$) corresponding to photo-dissociation regions (PDRs). The molecular content of the cloud decreases by 50\% because of the effect of radiation, even if dense molecular clumps ($n>10^4\,\cc$) are able to self-shield. At the end of this phase, the cloud presents an extended $\hi$ structure, whose collapse is prevented by the heating effect of non-ionizing radiation. Later, we identify a second phase in which dense clumps are consumed by rapid star formation (SFR$\simeq 0.1\,\msun\,\yr^{-1}$), so that radiation is free to escape and ionize the atomic gas in the computational box. 

After applying the correction for the formation of stars with mass $<1\,\msun$, we obtain that all the molecular gas in the cloud is exhausted in 1.6 Myr, and the final star formation efficiency (SFE) is 36 per cent. The short cloud lifetime that we found supports the picture of short-living GMCs \citep{Elmegreen2000}, which has also been supported by recent observations \citep{Chevance2019}. The obtained SFE efficiency is higher than the generally observed one, which is around 1-10\% \citep{Lada2010, Murray2010, Lee2016, Lada2016, Ochsendorf2017}, despite the accurate implementation of stellar feedback. In fact, the initial turbulence is rapidly dissipated in our simulations and the cloud collapses under its own self-gravity entailing an high star formation rate (${\rm SFR}_{\rm max}=0.1\,\msun\,\yr^{-1}$). The inclusion of turbulence driving, in order to effectively account for large-scale phenomena injecting kinetic energy in the ISM, would certainly help to sustain the turbulence level in the cloud, and hence decrease the SFE. We plan to add this feature to our simulations in a future work.

We have then analysed the radiation field inside the GMC, in the Habing and the ionizing bands. $G_0$ is rather homogeneous throughout the cloud, attaining a value around $10^{2}$ at the end of the simulation. On the other hand, the ionization parameter $U_{\rm ion}$ shows a broader distribution, and hence a larger spatial modulation inside the cloud. This is crucial for the interpretation of emission from galaxy simulations, where generally individual GMCs are not resolved, and therefore their properties are averaged. The distribution of the radiation field strongly affects line emission calculations, and the sub-grid patchiness of $U_{\rm ion}$ should be taken into account when computing the intensity of lines as [NII] and [OIII] \citep{Olsen2018, Pallottini2019}. Finally, we find that throughout the evolution the escape fraction of ionizing photons from the cloud is $f_{\rm ion, esc} \simlt 0.03$, also in agreement with similar studies \citep{Howard2018a, Kimm2019, He2020}.


\section*{Acknowledgments}

This work used the DiRAC@Durham facility managed by the Institute for Computational Cosmology on behalf of the STFC DiRAC HPC Facility (www.dirac.ac.uk). The equipment was funded by BEIS capital funding via STFC capital grants ST/K00042X/1, ST/P002293/1, ST/R002371/1 and ST/S002502/1, Durham University and STFC operations grant ST/R000832/1. DiRAC is part of the National e-Infrastructure.
AF and AL acknowledge support from the ERC Advanced Grant INTERSTELLAR H2020/740120.
Any dissemination of results must indicate that it reflects only the author’s view and that the Commission is not responsible for any use that may be made of the information it contains. 
Support from the Carl Friedrich von Siemens-Forschungspreis der Alexander von Humboldt-Stiftung Research Award is kindly acknowledged (AF).
MF acknowledges support from the European Research Council (ERC) under the European Union's Horizon 2020 research and innovation programme (grant agreement No 757535).
MF acknowledges support by Fondazione Cariplo, grant number 2018-2329.
We thank C. Federrath, T. Naab, and the participants of Sexten Workshop 2020: \quotes{The Interstellar Medium of High Redshift Galaxies} for fruitful discussions.
We acknowledge use of the Python programming language, Astropy \citep{AstropyCollaboration2013}, Matplotlib \citep{Hunter2007}, NumPy \citep{VanderWalt2011}, \pymses~\citep{Labadens2012}.

\section*{Data availability}
The data underlying this article will be shared on reasonable request to the corresponding author.


\bibliographystyle{style/mnras}
\bibliography{library}

\begin{thebibliography}{}
\makeatletter
\relax
\def\mn@urlcharsother{\let\do\@makeother \do\$\do\&\do\#\do\^\do\_\do\%\do\~}
\def\mn@doi{\begingroup\mn@urlcharsother \@ifnextchar [ {\mn@doi@}
  {\mn@doi@[]}}
\def\mn@doi@[#1]#2{\def\@tempa{#1}\ifx\@tempa\@empty \href
  {http://dx.doi.org/#2} {doi:#2}\else \href {http://dx.doi.org/#2} {#1}\fi
  \endgroup}
\def\mn@eprint#1#2{\mn@eprint@#1:#2::\@nil}
\def\mn@eprint@arXiv#1{\href {http://arxiv.org/abs/#1} {{\tt arXiv:#1}}}
\def\mn@eprint@dblp#1{\href {http://dblp.uni-trier.de/rec/bibtex/#1.xml}
  {dblp:#1}}
\def\mn@eprint@#1:#2:#3:#4\@nil{\def\@tempa {#1}\def\@tempb {#2}\def\@tempc
  {#3}\ifx \@tempc \@empty \let \@tempc \@tempb \let \@tempb \@tempa \fi \ifx
  \@tempb \@empty \def\@tempb {arXiv}\fi \@ifundefined
  {mn@eprint@\@tempb}{\@tempb:\@tempc}{\expandafter \expandafter \csname
  mn@eprint@\@tempb\endcsname \expandafter{\@tempc}}}

\bibitem[\protect\citeauthoryear{Alves, Lombardi  \& Lada}{Alves
  et~al.}{2007}]{Alves2007}
Alves J.,  Lombardi M.,   Lada C.~J.,  2007, \mn@doi [Astronomy \&
  Astrophysics] {10.1051/0004-6361:20066389}, 462, L17

\bibitem[\protect\citeauthoryear{Anderson, Bania, Jackson, Clemens, Heyer,
  Simon, Shah  \& Rathborne}{Anderson et~al.}{2009}]{Anderson2009}
Anderson L.~D.,  Bania T.~M.,  Jackson J.~M.,  Clemens D.~P.,  Heyer M.,  Simon
  R.,  Shah R.~Y.,   Rathborne J.~M.,  2009, \mn@doi [The Astrophysical Journal
  Supplement Series] {10.1088/0067-0049/181/1/255}, 181, 255

\bibitem[\protect\citeauthoryear{Andr{\'e} et~al.,}{Andr{\'e}
  et~al.}{2010}]{Andre2010}
Andr{\'e} P.,  et~al., 2010, \mn@doi [Astronomy \& Astrophysics]
  {10.1051/0004-6361/201014666}, 518, L102

\bibitem[\protect\citeauthoryear{{Astropy Collaboration} et~al.,}{{Astropy
  Collaboration} et~al.}{2013}]{AstropyCollaboration2013}
{Astropy Collaboration} et~al., 2013, \mn@doi [Astronomy and Astrophysics]
  {10.1051/0004-6361/201322068}, 558, A33

\bibitem[\protect\citeauthoryear{Beerer et~al.,}{Beerer
  et~al.}{2010}]{Beerer2010}
Beerer I.~M.,  et~al., 2010, \mn@doi [The Astrophysical Journal]
  {10.1088/0004-637X/720/1/679}, 720, 679

\bibitem[\protect\citeauthoryear{Bisbas, Unsch, Whitworth, Hubber  \&
  Walch}{Bisbas et~al.}{2011}]{Bisbas2011}
Bisbas T.~G.,  Unsch R.,  Whitworth A.~P.,  Hubber D.~A.,   Walch S.,  2011,
  \mn@doi [The Astrophysical Journal] {10.1088/0004-637X/736/2/142}, 736, 142

\bibitem[\protect\citeauthoryear{Bovino, Grassi, Capelo, Schleicher  \&
  Banerjee}{Bovino et~al.}{2016}]{Bovino2016}
Bovino S.,  Grassi T.,  Capelo P.~R.,  Schleicher D. R.~G.,   Banerjee R.,
  2016, \mn@doi [Astronomy \& Astrophysics] {10.1051/0004-6361/201628158}, A15,
  1

\bibitem[\protect\citeauthoryear{Bressan, Marigo, Girardi, Salasnich, Dal~Cero,
  Rubele  \& Nanni}{Bressan et~al.}{2012}]{Bressan2012}
Bressan A.,  Marigo P.,  Girardi L.,  Salasnich B.,  Dal~Cero C.,  Rubele S.,
  Nanni A.,  2012, \mn@doi [Monthly Notices of the Royal Astronomical Society]
  {10.1111/j.1365-2966.2012.21948.x}, 427, 127

\bibitem[\protect\citeauthoryear{Bron, Ag{\'u}ndez, Goicoechea  \&
  Cernicharo}{Bron et~al.}{2018}]{Bron2018}
Bron E.,  Ag{\'u}ndez M.,  Goicoechea J.~R.,   Cernicharo J.,  2018, preprint
  (arXiv:1801.01547)

\bibitem[\protect\citeauthoryear{Brunt, Heyer  \& Mac~Low}{Brunt
  et~al.}{2009}]{Brunt2009}
Brunt C.~M.,  Heyer M.~H.,   Mac~Low M.-M.,  2009, \mn@doi [A\&A]
  {10.1051/0004-6361/200911797}, 504, 883

\bibitem[\protect\citeauthoryear{Carpenter}{Carpenter}{2000}]{Carpenter2000}
Carpenter J.,  2000, \mn@doi [The Astronomical Journal]
  {https://iopscience.iop.org/article/10.1086/316845}, 120, 3139

\bibitem[\protect\citeauthoryear{Castelli \& Kurucz}{Castelli \&
  Kurucz}{2003}]{Castelli2003}
Castelli F.,  Kurucz R.~L.,  2003, Modelling of Stellar Atmospheres, 210, A20

\bibitem[\protect\citeauthoryear{Castor, Abbott  \& Klein}{Castor
  et~al.}{1975}]{Castor1975}
Castor J.~I.,  Abbott D.~C.,   Klein R.~I.,  1975, \mn@doi [ApJ]
  {10.1086/153315}, 195, 157

\bibitem[\protect\citeauthoryear{Cen}{Cen}{1992}]{Cen1992}
Cen R.,  1992, \mn@doi [The Astrophysical Journal Supplement Series]
  {10.1086/191630}, 78, 341

\bibitem[\protect\citeauthoryear{Ceverino \& Klypin}{Ceverino \&
  Klypin}{2009}]{Ceverino2009}
Ceverino D.,  Klypin A.,  2009, \mn@doi [The Astrophysical Journal]
  {10.1088/0004-637X/695/1/292}, 695, 292

\bibitem[\protect\citeauthoryear{Chandrasekhar}{Chandrasekhar}{1939}]{Chandrasekhar1939}
Chandrasekhar S.,  1939, Chicago

\bibitem[\protect\citeauthoryear{Chevance et~al.,}{Chevance
  et~al.}{2019}]{Chevance2019}
Chevance M.,  et~al., 2019, \mn@doi [Monthly Notices of the Royal Astronomical
  Society] {10.1093/mnras/stz3525}

\bibitem[\protect\citeauthoryear{Cioffi, McKee  \& Bertschinger}{Cioffi
  et~al.}{1988}]{Cioffi1988}
Cioffi D.~F.,  McKee C.~F.,   Bertschinger E.,  1988, \mn@doi [The
  Astrophysical Journal] {10.1086/166834}, 334, 252

\bibitem[\protect\citeauthoryear{Colombo et~al.,}{Colombo
  et~al.}{2014}]{Colombo2014}
Colombo D.,  et~al., 2014, \mn@doi [ApJ] {10.1088/0004-637X/784/1/3}, 784, 3

\bibitem[\protect\citeauthoryear{Dale, Bonnell, Clarke  \& Bate}{Dale
  et~al.}{2005}]{Dale2005}
Dale J.~E.,  Bonnell I.~A.,  Clarke C.~J.,   Bate M.~R.,  2005, \mn@doi
  [Monthly Notices of the Royal Astronomical Society]
  {10.1111/j.1365-2966.2005.08806.x}, 358, 291

\bibitem[\protect\citeauthoryear{Dale, Bonnell  \& Whitworth}{Dale
  et~al.}{2007a}]{Dale2007b}
Dale J.~E.,  Bonnell I.~A.,   Whitworth A.~P.,  2007a, \mn@doi [Monthly Notices
  of the Royal Astronomical Society] {10.1111/j.1365-2966.2006.11368.x}, 375,
  1291

\bibitem[\protect\citeauthoryear{Dale, Clark  \& Bonnell}{Dale
  et~al.}{2007b}]{Dale2007a}
Dale J.~E.,  Clark P.~C.,   Bonnell I.~A.,  2007b, \mn@doi [Monthly Notices of
  the Royal Astronomical Society] {10.1111/j.1365-2966.2007.11515.x}, 377, 535

\bibitem[\protect\citeauthoryear{Decataldo, Pallottini, Ferrara, Vallini  \&
  Gallerani}{Decataldo et~al.}{2019}]{Decataldo2019}
Decataldo D.,  Pallottini A.,  Ferrara A.,  Vallini L.,   Gallerani S.,  2019,
  \mn@doi [Monthly Notices of the Royal Astronomical Society]
  {10.1093/mnras/stz1527}

\bibitem[\protect\citeauthoryear{Deharveng, Lefloch, Massi, Brand, Kurtz,
  Zavagno  \& Caplan}{Deharveng et~al.}{2006}]{Deharveng2006}
Deharveng L.,  Lefloch B.,  Massi F.,  Brand J.,  Kurtz S.,  Zavagno A.,
  Caplan J.,  2006, \mn@doi [Astronomy and Astrophysics]
  {10.1051/0004-6361:20054641}, 458, 191

\bibitem[\protect\citeauthoryear{Deharveng et~al.,}{Deharveng
  et~al.}{2010}]{Deharveng2010}
Deharveng L.,  et~al., 2010, \mn@doi [Astronomy and Astrophysics]
  {10.1051/0004-6361/201014422}, 523, A6

\bibitem[\protect\citeauthoryear{Deparis, Aubert, Ocvirk, Chardin  \&
  Lewis}{Deparis et~al.}{2019}]{Deparis2019}
Deparis N.,  Aubert D.,  Ocvirk P.,  Chardin J.,   Lewis J.,  2019, \mn@doi
  [Astronomy and Astrophysics] {10.1051/0004-6361/201832889}, 622, A142

\bibitem[\protect\citeauthoryear{Dobbs}{Dobbs}{2015}]{Dobbs2015}
Dobbs C.~L.,  2015, \mn@doi [MNRAS] {10.1093/mnras/stu2585}, 447, 3390

\bibitem[\protect\citeauthoryear{Dwarkadas \& Gruszko}{Dwarkadas \&
  Gruszko}{2012}]{Dwarkadas2012}
Dwarkadas V.~V.,  Gruszko J.,  2012, \mn@doi [Monthly Notices of the Royal
  Astronomical Society] {10.1111/j.1365-2966.2011.19808.x}, 419, 1515

\bibitem[\protect\citeauthoryear{Elmegreen}{Elmegreen}{2000}]{Elmegreen2000}
Elmegreen B.~G.,  2000, \mn@doi [ApJ] {10.1086/308361}, 530, 277

\bibitem[\protect\citeauthoryear{Elmegreen}{Elmegreen}{2011a}]{Elmegreen2011}
Elmegreen B.~G.,  2011a, ] {10.1051/eas/1151004}, 51, 45

\bibitem[\protect\citeauthoryear{Elmegreen}{Elmegreen}{2011b}]{Elmegreen2011a}
Elmegreen B.~G.,  2011b, \mn@doi [The Astrophysical Journal]
  {10.1088/0004-637X/737/1/10}, 737, 10

\bibitem[\protect\citeauthoryear{Evans et~al.,}{Evans et~al.}{2009}]{Evans2009}
Evans N.~J.,  et~al., 2009, \mn@doi [The Astrophysical Journal Supplement
  Series] {10.1088/0067-0049/181/2/321}, 181, 321

\bibitem[\protect\citeauthoryear{Federrath \& Klessen}{Federrath \&
  Klessen}{2012}]{Federrath2012}
Federrath C.,  Klessen R.~S.,  2012, \mn@doi [The Astrophysical Journal]
  {10.1088/0004-637X/761/2/156}, 761, 156

\bibitem[\protect\citeauthoryear{Federrath \& Klessen}{Federrath \&
  Klessen}{2013}]{Federrath2013}
Federrath C.,  Klessen R.~S.,  2013, \mn@doi [The Astrophysical Journal]
  {10.1088/0004-637X/763/1/51}, 763, 51

\bibitem[\protect\citeauthoryear{Ferland, Korista, Verner, Ferguson, Kingdon
  \& Verner}{Ferland et~al.}{1998}]{Ferland1998}
Ferland G.~J.,  Korista K.~T.,  Verner D.~A.,  Ferguson J.~W.,  Kingdon J.~B.,
   Verner E.~M.,  1998, \mn@doi [Publications of the Astronomical Society of
  the Pacific] {10.1086/316190}, 110, 761

\bibitem[\protect\citeauthoryear{Ferrara, Vallini, Pallottini, Gallerani,
  Carniani, Kohandel, Decataldo  \& Behrens}{Ferrara
  et~al.}{2019}]{Ferrara2019}
Ferrara A.,  Vallini L.,  Pallottini A.,  Gallerani S.,  Carniani S.,  Kohandel
  M.,  Decataldo D.,   Behrens C.,  2019, \mn@doi [Monthly Notices of the Royal
  Astronomical Society] {10.1093/mnras/stz2031}, 489, 1

\bibitem[\protect\citeauthoryear{Garc{\'i}a, Bronfman, Nyman, Dame  \&
  Luna}{Garc{\'i}a et~al.}{2014}]{Garcia2014}
Garc{\'i}a P.,  Bronfman L.,  Nyman L.-{\AA}.,  Dame T.~M.,   Luna A.,  2014,
  \mn@doi [The Astrophysical Journal Supplement Series]
  {10.1088/0067-0049/212/1/2}, 212, 2

\bibitem[\protect\citeauthoryear{Gatto et~al.,}{Gatto et~al.}{2017}]{Gatto2017}
Gatto A.,  et~al., 2017, \mn@doi [Monthly Notices of the Royal Astronomical
  Society] {10.1093/mnras/stw3209}, 466, 1903

\bibitem[\protect\citeauthoryear{Geen, Hennebelle, Tremblin  \& Rosdahl}{Geen
  et~al.}{2016}]{Geen2016}
Geen S.,  Hennebelle P.,  Tremblin P.,   Rosdahl J.,  2016, \mn@doi [Monthly
  Notices of the Royal Astronomical Society] {10.1093/mnras/stw2235}, 463, 3129

\bibitem[\protect\citeauthoryear{Geen, Watson, Rosdahl, Bieri, Klessen  \&
  Hennebelle}{Geen et~al.}{2018}]{Geen2018}
Geen S.,  Watson S.~K.,  Rosdahl J.,  Bieri R.,  Klessen R.~S.,   Hennebelle
  P.,  2018, \mn@doi [Monthly Notices of the Royal Astronomical Society]
  {10.1093/mnras/sty2439}, 481, 2548

\bibitem[\protect\citeauthoryear{Glover \& Abel}{Glover \&
  Abel}{2008}]{Glover2008}
Glover S. C.~O.,  Abel T.,  2008, \mn@doi [Monthly Notices of the Royal
  Astronomical Society] {10.1111/j.1365-2966.2008.13224.x}, 388, 1627

\bibitem[\protect\citeauthoryear{Gnedin \& Hollon}{Gnedin \&
  Hollon}{2012}]{Gnedin2012}
Gnedin N.~Y.,  Hollon N.,  2012, \mn@doi [The Astrophysical Journal Supplement
  Series] {10.1088/0067-0049/202/2/13}, 202, 13

\bibitem[\protect\citeauthoryear{Grassi, Bovino, Schleicher, Prieto, Seifried,
  Simoncini  \& Gianturco}{Grassi et~al.}{2014}]{Grassi2014}
Grassi T.,  Bovino S.,  Schleicher D. R.~G.,  Prieto J.,  Seifried D.,
  Simoncini E.,   Gianturco F.~A.,  2014, \mn@doi [Monthly Notices of the Royal
  Astronomical Society] {10.1093/mnras/stu114}, 439, 2386

\bibitem[\protect\citeauthoryear{Grisdale, Agertz, Renaud  \& Romeo}{Grisdale
  et~al.}{2018}]{Grisdale2018}
Grisdale K.,  Agertz O.,  Renaud F.,   Romeo A.~B.,  2018, \mn@doi [Monthly
  Notices of the Royal Astronomical Society] {10.1093/mnras/sty1595}, 479, 3167

\bibitem[\protect\citeauthoryear{Gritschneder, Naab, Burkert, Walch, Heitsch
  \& Wetzstein}{Gritschneder et~al.}{2009}]{Gritschneder2009}
Gritschneder M.,  Naab T.,  Burkert A.,  Walch S.,  Heitsch F.,   Wetzstein M.,
   2009, \mn@doi [Monthly Notices of the Royal Astronomical Society]
  {10.1111/j.1365-2966.2008.14185.x}, 393, 21

\bibitem[\protect\citeauthoryear{Haardt \& Madau}{Haardt \&
  Madau}{2012}]{Haardt2012}
Haardt F.,  Madau P.,  2012, \mn@doi [The Astrophysical Journal]
  {10.1088/0004-637X/746/2/125}, 746, 125

\bibitem[\protect\citeauthoryear{Haid, Walch, Seifried, W{\"u}nsch, Dinnbier
  \& Naab}{Haid et~al.}{2018}]{Haid2018}
Haid S.,  Walch S.,  Seifried D.,  W{\"u}nsch R.,  Dinnbier F.,   Naab T.,
  2018, \mn@doi [Monthly Notices of the Royal Astronomical Society]
  {10.1093/mnras/sty1315}, 478, 4799

\bibitem[\protect\citeauthoryear{Haid, Walch, Seifried, W{\"u}nsch, Dinnbier
  \& Naab}{Haid et~al.}{2019}]{Haid2019}
Haid S.,  Walch S.,  Seifried D.,  W{\"u}nsch R.,  Dinnbier F.,   Naab T.,
  2019, \mn@doi [Monthly Notices of the Royal Astronomical Society]
  {10.1093/mnras/sty2938}, 482, 4062

\bibitem[\protect\citeauthoryear{Hartmann, {Ballesteros-Paredes}  \&
  Bergin}{Hartmann et~al.}{2001}]{Hartmann2001}
Hartmann L.,  {Ballesteros-Paredes} J.,   Bergin E.~A.,  2001, \mn@doi [ApJ]
  {10.1086/323863}, 562, 852

\bibitem[\protect\citeauthoryear{He, Ricotti  \& Geen}{He
  et~al.}{2019}]{He2019}
He C.-C.,  Ricotti M.,   Geen S.,  2019, preprint (arXiv:1904.07889)

\bibitem[\protect\citeauthoryear{He, Ricotti  \& Geen}{He
  et~al.}{2020}]{He2020}
He C.-C.,  Ricotti M.,   Geen S.,  2020, \mn@doi [Monthly Notices of the Royal
  Astronomical Society] {10.1093/mnras/staa165}, 492, 4858

\bibitem[\protect\citeauthoryear{Heyer, Krawczyk, Duval  \& Jackson}{Heyer
  et~al.}{2009}]{Heyer2009}
Heyer M.,  Krawczyk C.,  Duval J.,   Jackson J.~M.,  2009, \mn@doi [The
  Astrophysical Journal] {10.1088/0004-637X/699/2/1092}, 699, 1092

\bibitem[\protect\citeauthoryear{Hollenbach \& McKee}{Hollenbach \&
  McKee}{1979}]{Hollenbach1979}
Hollenbach D.,  McKee C.~F.,  1979, \mn@doi [The Astrophysical Journal
  Supplement Series] {10.1086/190631}, 41, 555

\bibitem[\protect\citeauthoryear{Hosokawa \& Inutsuka}{Hosokawa \&
  Inutsuka}{2005}]{Hosokawa2005}
Hosokawa T.,  Inutsuka S.-i.,  2005, \mn@doi [The Astrophysical Journal]
  {10.1086/428648}, 623, 917

\bibitem[\protect\citeauthoryear{Hosokawa \& Inutsuka}{Hosokawa \&
  Inutsuka}{2006a}]{Hosokawa2006}
Hosokawa T.,  Inutsuka S.-i.,  2006a, \mn@doi [The Astrophysical Journal]
  {10.1086/504789}, 646, 240

\bibitem[\protect\citeauthoryear{Hosokawa \& Inutsuka}{Hosokawa \&
  Inutsuka}{2006b}]{Hosokawa2006a}
Hosokawa T.,  Inutsuka S.-i.,  2006b, \mn@doi [The Astrophysical Journal
  Letters] {10.1086/507887}, 648, L131

\bibitem[\protect\citeauthoryear{Howard, Pudritz  \& Harris}{Howard
  et~al.}{2017}]{Howard2017}
Howard C.~S.,  Pudritz R.~E.,   Harris W.~E.,  2017, \mn@doi [Monthly Notices
  of the Royal Astronomical Society] {10.1093/mnras/stx1363}, 470, 3346

\bibitem[\protect\citeauthoryear{Howard, Pudritz, Harris  \& Klessen}{Howard
  et~al.}{2018}]{Howard2018a}
Howard C.~S.,  Pudritz R.~E.,  Harris W.~E.,   Klessen R.~S.,  2018, \mn@doi
  [Monthly Notices of the Royal Astronomical Society] {10.1093/mnras/stx3276},
  475, 3121

\bibitem[\protect\citeauthoryear{Hughes et~al.,}{Hughes
  et~al.}{2013}]{Hughes2013}
Hughes A.,  et~al., 2013, \mn@doi [ApJ] {10.1088/0004-637X/779/1/46}, 779, 46

\bibitem[\protect\citeauthoryear{Hunter}{Hunter}{2007}]{Hunter2007}
Hunter J.~D.,  2007, \mn@doi [Computing in Science Engineering]
  {10.1109/MCSE.2007.55}, 9, 90

\bibitem[\protect\citeauthoryear{Jura}{Jura}{1975}]{Jura1975}
Jura M.,  1975, \mn@doi [The Astrophysical Journal] {10.1086/153545}, 197, 575

\bibitem[\protect\citeauthoryear{Kahn}{Kahn}{1954}]{Kahn1954}
Kahn F.~D.,  1954, Bulletin of the Astronomical Institutes of the Netherlands,
  12, 187

\bibitem[\protect\citeauthoryear{Kaufman, Wolfire, Hollenbach  \&
  Luhman}{Kaufman et~al.}{1999}]{Kaufman1999}
Kaufman M.~J.,  Wolfire M.~G.,  Hollenbach D.~J.,   Luhman M.~L.,  1999,
  \mn@doi [The Astrophysical Journal] {10.1086/308102}, 527, 795

\bibitem[\protect\citeauthoryear{Kennicutt}{Kennicutt}{1998}]{Kennicutt1998}
Kennicutt R.~C.,  1998, \mn@doi [Annual Review of Astronomy and Astrophysics]
  {10.1146/annurev.astro.36.1.189}, 36, 189

\bibitem[\protect\citeauthoryear{{Kessel-Deynet} \& Burkert}{{Kessel-Deynet} \&
  Burkert}{2003}]{Kessel-Deynet2003}
{Kessel-Deynet} O.,  Burkert A.,  2003, \mn@doi [Monthly Notices of the Royal
  Astronomical Society] {10.1046/j.1365-8711.2003.05737.x}, 338, 545

\bibitem[\protect\citeauthoryear{Kim, Kim  \& Ostriker}{Kim
  et~al.}{2018}]{Kim2018}
Kim J.-G.,  Kim W.-T.,   Ostriker E.~C.,  2018, \mn@doi [The Astrophysical
  Journal] {10.3847/1538-4357/aabe27}, 859, 68

\bibitem[\protect\citeauthoryear{Kimm, Blaizot, Garel, {Michel-Dansac}, Katz,
  Rosdahl, Verhamme  \& Haehnelt}{Kimm et~al.}{2019}]{Kimm2019}
Kimm T.,  Blaizot J.,  Garel T.,  {Michel-Dansac} L.,  Katz H.,  Rosdahl J.,
  Verhamme A.,   Haehnelt M.,  2019, \mn@doi [Monthly Notices of the Royal
  Astronomical Society] {10.1093/mnras/stz989}, 486, 2215

\bibitem[\protect\citeauthoryear{Kippenhahn \& Weigert}{Kippenhahn \&
  Weigert}{1990}]{kippenhahn1990}
Kippenhahn R.,  Weigert A.,  1990, Stellar Structure and Evolution, XVI, 468
  pp. 192 figs.. Springer-Verlag Berlin Heidelberg New York. Also Astronomy and
  Astrophysics Library

\bibitem[\protect\citeauthoryear{Kroupa}{Kroupa}{2001}]{Kroupa2001}
Kroupa P.,  2001, \mn@doi [Monthly Notices of the Royal Astronomical Society]
  {10.1046/j.1365-8711.2001.04022.x}, 322, 231

\bibitem[\protect\citeauthoryear{Kruijssen et~al.,}{Kruijssen
  et~al.}{2019}]{Kruijssen2019}
Kruijssen J. M.~D.,  et~al., 2019, \mn@doi [Nature]
  {10.1038/s41586-019-1194-3}, 569, 519

\bibitem[\protect\citeauthoryear{Krumholz \& McKee}{Krumholz \&
  McKee}{2005}]{Krumholz2005a}
Krumholz M.~R.,  McKee C.~F.,  2005, \mn@doi [The Astrophysical Journal]
  {10.1086/431734}, 630, 250

\bibitem[\protect\citeauthoryear{Krumholz, Dekel  \& McKee}{Krumholz
  et~al.}{2012}]{Krumholz2012}
Krumholz M.~R.,  Dekel A.,   McKee C.~F.,  2012, \mn@doi [The Astrophysical
  Journal] {10.1088/0004-637X/745/1/69}, 745, 69

\bibitem[\protect\citeauthoryear{Labadens, Chapon, Pomar{\'e}de  \&
  Teyssier}{Labadens et~al.}{2012}]{Labadens2012}
Labadens M.,  Chapon D.,  Pomar{\'e}de D.,   Teyssier R.,  2012, in
  Astronomical {{Data Analysis Software}} and {{Systems XXI}}. p.~837

\bibitem[\protect\citeauthoryear{Lada}{Lada}{2016}]{Lada2016}
Lada C.~J.,  2016, ] {10.1017/S1743921315005955}, 314, 8

\bibitem[\protect\citeauthoryear{Lada, Alves  \& Lada}{Lada
  et~al.}{1999}]{Lada1999}
Lada C.~J.,  Alves J.,   Lada E.~A.,  1999, \mn@doi [The Astrophysical Journal]
  {10.1086/306756}, 512, 250

\bibitem[\protect\citeauthoryear{Lada, Lombardi  \& Alves}{Lada
  et~al.}{2010}]{Lada2010}
Lada C.~J.,  Lombardi M.,   Alves J.~F.,  2010, \mn@doi [The Astrophysical
  Journal] {10.1088/0004-637X/724/1/687}, 724, 687

\bibitem[\protect\citeauthoryear{Larson}{Larson}{1981}]{Larson1981}
Larson R.~B.,  1981, \mn@doi [Monthly Notices of the Royal Astronomical
  Society] {10.1093/mnras/194.4.809}, 194, 809

\bibitem[\protect\citeauthoryear{Le~Petit, Nehm{\'e}, Le~Bourlot  \&
  Roueff}{Le~Petit et~al.}{2006}]{LePetit2006}
Le~Petit F.,  Nehm{\'e} C.,  Le~Bourlot J.,   Roueff E.,  2006, \mn@doi [The
  Astrophysical Journal Supplement Series] {10.1086/503252}, 164, 506

\bibitem[\protect\citeauthoryear{Lee, {Miville-Desch{\^e}nes}  \& Murray}{Lee
  et~al.}{2016}]{Lee2016}
Lee E.~J.,  {Miville-Desch{\^e}nes} M.-A.,   Murray N.~W.,  2016, \mn@doi [The
  Astrophysical Journal] {10.3847/1538-4357/833/2/229}, 833, 229

\bibitem[\protect\citeauthoryear{Leitherer, Robert  \& Drissen}{Leitherer
  et~al.}{1992}]{Leitherer1992}
Leitherer C.,  Robert C.,   Drissen L.,  1992, \mn@doi [The Astrophysical
  Journal] {10.1086/172089}, 401, 596

\bibitem[\protect\citeauthoryear{Leung et~al.,}{Leung et~al.}{2019}]{Leung2019}
Leung T. K.~D.,  et~al., 2019, \mn@doi [The Astrophysical Journal]
  {10.3847/1538-4357/aaf860}, 871, 85

\bibitem[\protect\citeauthoryear{Liu et~al.,}{Liu et~al.}{2016}]{Liu2016}
Liu H.-L.,  et~al., 2016, \mn@doi [The Astrophysical Journal]
  {10.3847/0004-637X/818/1/95}, 818, 95

\bibitem[\protect\citeauthoryear{Matzner}{Matzner}{2002}]{Matzner2002}
Matzner C.~D.,  2002, \mn@doi [The Astrophysical Journal] {10.1086/338030},
  566, 302

\bibitem[\protect\citeauthoryear{McKee \& Cowie}{McKee \&
  Cowie}{1977}]{McKee1977}
McKee C.~F.,  Cowie L.~L.,  1977, \mn@doi [The Astrophysical Journal]
  {10.1086/155350}, 215, 213

\bibitem[\protect\citeauthoryear{Murray, Quataert  \& Thompson}{Murray
  et~al.}{2010}]{Murray2010}
Murray N.,  Quataert E.,   Thompson T.~A.,  2010, \mn@doi [The Astrophysical
  Journal] {10.1088/0004-637X/709/1/191}, 709, 191

\bibitem[\protect\citeauthoryear{Ochsendorf, Meixner, {Roman-Duval}, Rahman  \&
  Evans}{Ochsendorf et~al.}{2017}]{Ochsendorf2017}
Ochsendorf B.~B.,  Meixner M.,  {Roman-Duval} J.,  Rahman M.,   Evans N.~J.,
  2017, \mn@doi [The Astrophysical Journal] {10.3847/1538-4357/aa704a}, 841,
  109

\bibitem[\protect\citeauthoryear{Ocvirk, Aubert, Chardin, Deparis  \&
  Lewis}{Ocvirk et~al.}{2019}]{Ocvirk2019}
Ocvirk P.,  Aubert D.,  Chardin J.,  Deparis N.,   Lewis J.,  2019, \mn@doi
  [Astronomy \& Astrophysics] {10.1051/0004-6361/201832923}, 626, A77

\bibitem[\protect\citeauthoryear{Olsen et~al.,}{Olsen et~al.}{2018}]{Olsen2018}
Olsen K.,  et~al., 2018, \mn@doi [Galaxies] {10.3390/galaxies6040100}, 6, 100

\bibitem[\protect\citeauthoryear{Ostriker \& McKee}{Ostriker \&
  McKee}{1988}]{Ostriker1988}
Ostriker J.~P.,  McKee C.~F.,  1988, Reviews of Modern Physics, 60, 1

\bibitem[\protect\citeauthoryear{Padoan, Haugb{\o}lle  \& Nordlund}{Padoan
  et~al.}{2012}]{Padoan2012}
Padoan P.,  Haugb{\o}lle T.,   Nordlund {\AA}.,  2012, \mn@doi [The
  Astrophysical Journal Letters] {10.1088/2041-8205/759/2/L27}, 759, L27

\bibitem[\protect\citeauthoryear{Pallottini et~al.,}{Pallottini
  et~al.}{2019}]{Pallottini2019}
Pallottini A.,  et~al., 2019, \mn@doi [Monthly Notices of the Royal
  Astronomical Society] {10.1093/mnras/stz1383}, 487, 1689

\bibitem[\protect\citeauthoryear{Raskutti, Ostriker  \& Skinner}{Raskutti
  et~al.}{2016}]{Raskutti2016}
Raskutti S.,  Ostriker E.~C.,   Skinner M.~A.,  2016, \mn@doi [The
  Astrophysical Journal] {10.3847/0004-637X/829/2/130}, 829, 130

\bibitem[\protect\citeauthoryear{Richings, Schaye  \& Oppenheimer}{Richings
  et~al.}{2014}]{Richings2014b}
Richings A.~J.,  Schaye J.,   Oppenheimer B.~D.,  2014, \mn@doi [Monthly
  Notices of the Royal Astronomical Society] {10.1093/mnras/stu1046}, 442, 2780

\bibitem[\protect\citeauthoryear{Rosdahl, Blaizot, Aubert, Stranex  \&
  Teyssier}{Rosdahl et~al.}{2013}]{Rosdahl2013}
Rosdahl J.,  Blaizot J.,  Aubert D.,  Stranex T.,   Teyssier R.,  2013, \mn@doi
  [Monthly Notices of the Royal Astronomical Society] {10.1093/mnras/stt1722},
  436, 2188

\bibitem[\protect\citeauthoryear{Rosdahl et~al.,}{Rosdahl
  et~al.}{2018}]{Rosdahl2018}
Rosdahl J.,  et~al., 2018, \mn@doi [Monthly Notices of the Royal Astronomical
  Society] {10.1093/mnras/sty1655}, 479, 994

\bibitem[\protect\citeauthoryear{Schmidt}{Schmidt}{1959}]{Schmidt1959}
Schmidt M.,  1959, \mn@doi [The Astrophysical Journal] {10.1086/146614}, 129,
  243

\bibitem[\protect\citeauthoryear{Sedov}{Sedov}{1958}]{Sedov1958}
Sedov L.~I.,  1958, \mn@doi [RvMP] {10.1103/RevModPhys.30.1077}, 30, 1077

\bibitem[\protect\citeauthoryear{Smartt}{Smartt}{2009}]{Smartt2009}
Smartt S.~J.,  2009, \mn@doi [Annual Review of Astronomy and Astrophysics]
  {10.1146/annurev-astro-082708-101737}, 47, 63

\bibitem[\protect\citeauthoryear{Sternberg \& Dalgarno}{Sternberg \&
  Dalgarno}{1989}]{Sternberg1989}
Sternberg A.,  Dalgarno A.,  1989, The Astrophysical Journal, 338, 197

\bibitem[\protect\citeauthoryear{Str{\"o}mgren}{Str{\"o}mgren}{1939}]{Stromgren1939}
Str{\"o}mgren B.,  1939, \mn@doi [The Astrophysical Journal] {10.1086/144074},
  89, 526

\bibitem[\protect\citeauthoryear{Teyssier}{Teyssier}{2002}]{Teyssier2002}
Teyssier R.,  2002, \mn@doi [Astronomy \& Astrophysics] {10.1051/0004-6361},
  385, 337

\bibitem[\protect\citeauthoryear{Thornton, Gaudlitz, Janka  \&
  Steinmetz}{Thornton et~al.}{1998}]{Thornton1998}
Thornton K.,  Gaudlitz M.,  Janka H.-T.,   Steinmetz M.,  1998, \mn@doi [The
  Astrophysical Journal] {10.1086/305704}, 500, 95

\bibitem[\protect\citeauthoryear{Tielens \& Hollenbach}{Tielens \&
  Hollenbach}{1985}]{Tielens1985}
Tielens A. G. G.~M.,  Hollenbach D.~J.,  1985, \mn@doi [The Astrophysical
  Journal] {10.1086/163111}, 291, 722

\bibitem[\protect\citeauthoryear{Vallini, Ferrara, Pallottini  \&
  Gallerani}{Vallini et~al.}{2017}]{Vallini2017}
Vallini L.,  Ferrara A.,  Pallottini A.,   Gallerani S.,  2017, \mn@doi
  [Monthly Notices of the Royal Astronomical Society] {10.1093/mnras/stx180},
  467, 1300

\bibitem[\protect\citeauthoryear{Vallini, Pallottini, Ferrara, Gallerani,
  Sobacchi  \& Behrens}{Vallini et~al.}{2018}]{Vallini2018}
Vallini L.,  Pallottini A.,  Ferrara A.,  Gallerani S.,  Sobacchi E.,   Behrens
  C.,  2018, \mn@doi [Monthly Notices of the Royal Astronomical Society]
  {10.1093/mnras/stx2376}, 473, 271

\bibitem[\protect\citeauthoryear{{Van der Walt}, Colbert  \& Varoquaux}{{Van
  der Walt} et~al.}{2011}]{VanderWalt2011}
{Van der Walt} S.,  Colbert S.~C.,   Varoquaux G.,  2011, \mn@doi [Computing in
  Science Engineering] {10.1109/MCSE.2011.37}, 13, 22

\bibitem[\protect\citeauthoryear{{V{\'a}zquez-Semadeni}, Col{\'i}n, G{\'o}mez,
  {Ballesteros-Paredes}  \& Watson}{{V{\'a}zquez-Semadeni}
  et~al.}{2010}]{Vazquez-Semadeni2010}
{V{\'a}zquez-Semadeni} E.,  Col{\'i}n P.,  G{\'o}mez G.~C.,
  {Ballesteros-Paredes} J.,   Watson A.~W.,  2010, \mn@doi [The Astrophysical
  Journal] {10.1088/0004-637X/715/2/1302}, 715, 1302

\bibitem[\protect\citeauthoryear{Walch \& Naab}{Walch \&
  Naab}{2015}]{Walch2015b}
Walch S.,  Naab T.,  2015, \mn@doi [Monthly Notices of the Royal Astronomical
  Society] {10.1093/mnras/stv1155}, 451, 2757

\bibitem[\protect\citeauthoryear{Walch, Whitworth, Bisbas, W{\"u}nsch  \&
  Hubber}{Walch et~al.}{2012}]{Walch2012}
Walch S.~K.,  Whitworth A.~P.,  Bisbas T.,  W{\"u}nsch R.,   Hubber D.,  2012,
  \mn@doi [Monthly Notices of the Royal Astronomical Society]
  {10.1111/j.1365-2966.2012.21767.x}, 427, 625

\bibitem[\protect\citeauthoryear{Walch, Whitworth, Bisbas  \& Hubber}{Walch
  et~al.}{2013}]{Walch2013}
Walch S.,  Whitworth A.~P.,  Bisbas T.~G.,   Hubber D.~A.,  2013, \mn@doi
  [Monthly Notices of the Royal Astronomical Society] {10.1093/mnras/stt1115},
  435, 917

\bibitem[\protect\citeauthoryear{Walch et~al.,}{Walch et~al.}{2015}]{Walch2015}
Walch S.,  et~al., 2015, \mn@doi [Monthly Notices of the Royal Astronomical
  Society] {10.1093/mnras/stv1975}, 454, 238

\bibitem[\protect\citeauthoryear{Weaver, McCray, Castor, Shapiro  \&
  Castor}{Weaver et~al.}{1977}]{Weaver1977}
Weaver R.,  McCray R.,  Castor J.,  Shapiro P.,   Castor J.,  1977, \mn@doi
  [The Astrophysical Journal] {10.1086/181908}, 218, 377

\bibitem[\protect\citeauthoryear{Weingartner \& Draine}{Weingartner \&
  Draine}{2001}]{Weingartner2001}
Weingartner J.~C.,  Draine B.~T.,  2001, \mn@doi [The Astrophysical Journal]
  {10.1086/318651}, 548, 296

\bibitem[\protect\citeauthoryear{Whitworth}{Whitworth}{1979}]{Whitworth1979}
Whitworth A.,  1979, \mn@doi [Monthly Notices of the Royal Astronomical
  Society] {10.1093/mnras/186.1.59}, 186, 59

\bibitem[\protect\citeauthoryear{Williams \& McKee}{Williams \&
  McKee}{1997}]{Williams1997}
Williams J.~P.,  McKee C.~F.,  1997, \mn@doi [The Astrophysical Journal]
  {10.1086/303588}, 476, 166

\bibitem[\protect\citeauthoryear{Williams, Bisbas, Haworth  \& Mackey}{Williams
  et~al.}{2018}]{Williams2018}
Williams R. J.~R.,  Bisbas T.~G.,  Haworth T.~J.,   Mackey J.,  2018, \mn@doi
  [Monthly Notices of the Royal Astronomical Society] {10.1093/mnras/sty1484},
  479, 2016

\bibitem[\protect\citeauthoryear{Yep \& White}{Yep \& White}{2020}]{Yep2020}
Yep A.~C.,  White R.~J.,  2020, \mn@doi [The Astrophysical Journal]
  {10.3847/1538-4357/ab6333}, 889, 50

\bibitem[\protect\citeauthoryear{Zavagno, Deharveng, Comer{\'o}n, Brand, Massi,
  Caplan  \& Russeil}{Zavagno et~al.}{2006}]{Zavagno2006}
Zavagno A.,  Deharveng L.,  Comer{\'o}n F.,  Brand J.,  Massi F.,  Caplan J.,
  Russeil D.,  2006, \mn@doi [Astronomy and Astrophysics]
  {10.1051/0004-6361:20053952}, 446, 171

\makeatother
\end{thebibliography}


\FloatBarrier 
\appendix

\section{Tests of the wind model}

\begin{figure}
\centering
\includegraphics[width=0.5\textwidth]{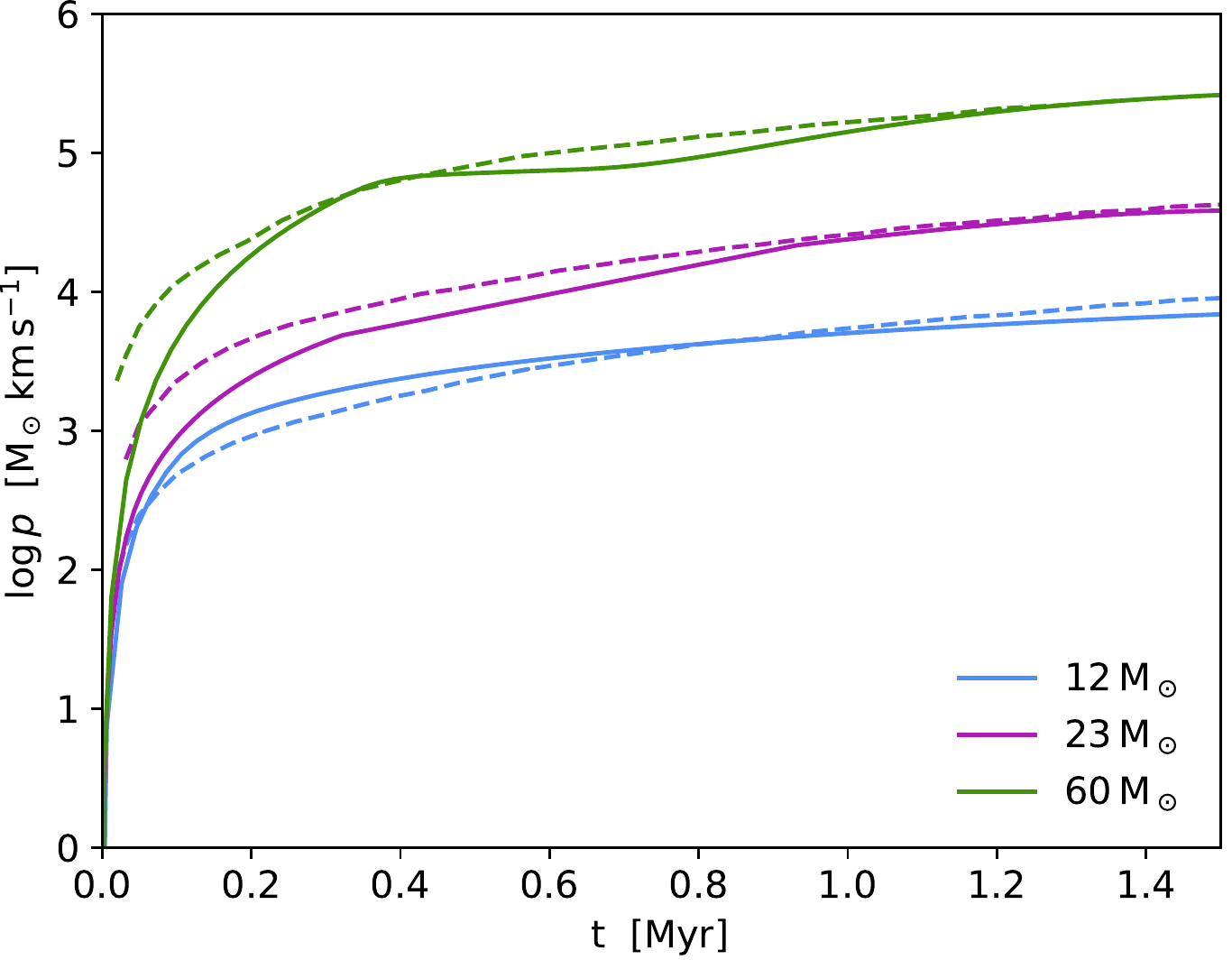}
\caption {Momentum injected in the ISM by stars with different masses (12 $\msun$, 23 $\msun$ and 60 $\msun$), placed at the center of a uniform density ($n=100\,\cc$, $T=10$ K). The solid and dashed lines represent the results of our tests and \citet{Haid2018} tests, showing a very good agreement. 
\label{wind_test}
}
\end{figure}

In our implementation of stellar winds, mass, momentum and kinetic energy are injected in the 27 cells neighbouring the position of the stellar particle, as detailed in Sec. \ref{sec:sub:model_winds}. This implementation is similar to the one in \citet{Gatto2017} and \citet[][H18 from here on]{Haid2018} simulations, with the only difference that they perform the injection over a sphere with the radius of 4 cells at the maximum refinement level. We run 3 tests to compare the momentum injection in the ISM. A star with mass 12 $\msun$, 23 $\msun$ or 60 $\msun$ is placed at the center of a uniform box with size 20 pc, filled with uniform medium with density $n=100\,\cc$ and initial temperature $T=10$ K. The same tests have been run in H18 (in particular, their CNM case), and this allows for a direct comparison. In particular, we compare the total momentum injection (including both wind and radiation) in the surrounding medium. The results are shown in Fig. \ref{wind_test} (cfr. with their Fig. 4). The three colours stand for different stellar masses, and the results of our tests and H18 tests are shown respectively with solid and dashed lines. For all the test cases, we find a very good agreement with H18 at large times. A small discrepancy, less than 10\%, can be noticed only at early times ($t < 0.5$ Myr), and it can be attributed to (1) the different cooling model adopted and (2) the fact that we adopt a constant momentum injection from the wind (while H18 accounts for stellar evolution).

\label{lastpage}
\end{document}